\documentclass{article}
\usepackage{fullpage}
\usepackage[utf8]{inputenc}
\usepackage{amsmath}
\usepackage{amssymb} 
\usepackage{upgreek}
\usepackage{mathtools}
\usepackage{graphicx}
\usepackage{dsfont}
\usepackage{listings}
  \usepackage{xr}
\usepackage{verbatim }
\usepackage{filecontents}
\usepackage{bm}
\usepackage[labelfont=bf]{caption}
\usepackage{subcaption}
\usepackage{comment}

\usepackage[hidelinks]{hyperref}
\usepackage{natbib}
\usepackage{multirow}
\newcommand{\animage}[2]{\includegraphics[width=#1\linewidth]{#2}}

\DeclareMathOperator*{\argmin}{arg\,min}
\def\TA{\textsc{TA}}

\usepackage{amsthm}

\theoremstyle{definition}

\newtheorem{model}{Model}[subsection]

\newcommand{\footremember}[2]{%
	\footnote{#2}
	\newcounter{#1}
	\setcounter{#1}{\value{footnote}}%
}
\newcommand{\footrecall}[1]{%
	\footnotemark[\value{#1}]%
} 
\newcommand{\appendixnumberline}[1]{Appendix\space}

\let\oldappendix\appendix
\makeatletter
\renewcommand{\appendix}{%
	\addtocontents{toc}{\let\protect\numberline\protect\appendixnumberline}%
	\renewcommand{\@seccntformat}[1]{Appendix~\csname the##1\endcsname\quad}%
	\oldappendix
}
\makeatother

\title{Targeting functional parameters with semiparametric Bayesian inference}
\author{%
	Vivian Y. Meng\footremember{mcgill}{Department of Mathematics and Statistics, McGill University, Canada}%
	\and David A. Stephens\footrecall{mcgill}%
}
\date{}

\begin{document}
	\makeatletter\@input{xx.tex}\makeatother
	\maketitle
	\begin{abstract}
		Typical Bayesian inference requires parameter identification via likelihood parameterization, which has invited criticism for being less flexible than the Frequentist framework and subject to misspecification. Though misspecification may be avoided by functional parameter inference under a nonparametric model space, there does not exist a flexible Bayesian semiparametric model that would allow full control over the marginal prior over any general functional parameter. We present the technique of $\theta$-augmentation which helps us manipulate nonparametric models into semiparametric ones that directly target any functional parameter. The method allows Bayesian probabilistic statements to be drawn for any estimator that is defined as a functional of the empirical distribution without requiring a likelihood function, thus providing a path to Bayesian analysis in problems like causal inference and censoring where there do not exist well-accepted likelihood functions.
		
		\textit{Keywords:} Bayesian; Functional parameter; Semiparametric.
	\end{abstract}

	\section{Introduction}
	This paper is concerned with the problem of making Bayesian inference regarding a functional parameter of the data-generating distribution of an independent and identically distributed collection of random variables.  Specifically, we construct inference for $\theta: \mathcal{F} \rightarrow \Theta$, a finite-dimensional map from the space of distributions, $\mathcal{F}$, to the parameter space, $\Theta$, while retaining a marginal subjective prior distribution $p_\theta$ for the functional of interest. Functionals regularly targeted in the literature include the minimizer of an expected loss function, that is, $\argmin_{t \in \Theta} \int l(x, t) F_0(\mathrm{d}x)$, solution to estimating equations, that is, ${\lbrace t\in \Theta: \int g(x,t)F_0(\mathrm{d}x)=0\rbrace}$, or in general, any function of the data-generating distribution, $F_0$. The approach of targeting functional parameters is uncommon in Bayesian inference and differs from the usual goal of targeting parameters that appear naturally in, say, de Finetti's representation, or in a likelihood formulation. We believe that targeting such functional parameters is key to avoiding model misspecification in general, and provides proper Bayesian inference in problems without well-accepted parametric likelihood functions.
	

	
	In typical Bayesian inference, we have, through Bayes rule, an explicit expression for the posterior distribution of likelihood parameters that is proportional to the likelihood function times a prior distribution. Under the most general form of de Finetti's representation theorem for infinitely exchangeable sequence of observables, the likelihood function is based on a nonparametric distribution function, and the corresponding prior, which we denote by $Q$, is a distribution over the space of nonparametric distribution functions. Additional restrictions on the observable, e.g. symmetry (see \cite{bernardo1994bayesian}, Chapter 4), leads to a reduction of the support of prior $Q$, to concentrate over the space of parametric models. These additional restrictions are not always correct, in the sense of the true data-generating model being in the support of $Q$, but make Bayesian calculations relatively simple. Furthermore, if we choose parametric likelihoods where a subset of the parameter vector coincides with the target parameter, then our subjective prior $p_\theta$ can be used in the definition of $Q$ thus ensuring coherence between our marginal subjective prior and the full Bayesian prior model. Due to these advantages, an overwhelming number of applications of Bayesian inference are restricted to parametric models with the target parameter coinciding with a part of the likelihood parameterization. However, since parametric likelihoods arise out of extraneous restrictions beyond exchangeability, they are subject to misspecification \citep{walker2013bayesian}. Some flexibility is gained by using semiparametric likelihoods, but the target parameter typically remains defined through the parameterization of a parametric conditional likelihood which facilitates marginal prior specification but is subject to misspecification, e.g. \cite{ray2020semiparametric}. Furthermore, there are simply problems where we have trouble specify a likelihood that is parameterized in terms of the target of inference, e.g. causal inference (see \cite{Stephens22-BA1322}). The practice of defining the target as a subset of the likelihood parameter is therefore thought of as a limitation of Bayesian inference compared to Frequentist methods where plug-in methods relying on the empirical distribution function essentially allow the targeting of any parameter. This limitation together with the issue of misspecification are not features of the Bayesian paradigm, but rather consequences of the simplifications made for the sake of more straightforward computation; yet they have led many researchers to shy away from the likelihood altogether.
	
	The desire to be likelihood-free has led to the advent of pseudo-Bayesian methods for likelihood-free inference. One method, the Bayesian empirical likelihood (BEL) method \citep{lazar03}, uses the empirical likelihood function \citep{owen1990empirical} for a parameter defined as a functional of the empirical distribution instead of the true likelihood function, in a Bayesian style prior-to-posterior update. The obvious downside is that the method is not genuinely Bayesian, as the empirical likelihood function is data-dependent. In problems where the definition of the target parameter requires intermediary nuisance parameters, whether we profile over the nuisance parameter first in the profile empirical likelihood function or perform a marginalization of the joint BEL posterior, the choice is arbitrary and leads to different posterior distributions. In the general Bayes (GB) method \citep{bissiri2016general}, which coincides with the Gibbs posterior \citep{jiang2008gibbs, zhang2006e}, the target parameter is defined and connected to the observable random quantity through a loss function rather than the likelihood. Based on a genuine loss function of the form $l(x,\theta)$, where $x$ is the observed data and $\theta$ the target parameter, the GB solution, which is the Gibbs posterior, is simply proportional to $\exp(-l(x,\theta))p_\theta(\theta)$ with our subjective prior given by $p_\theta$. The GB methods does have a criterion for parameter identification, that is, the requirement that $\argmin_{\theta \in \Theta} \int l(x,\theta) \mathrm{d}F_0(x)$ identifies the true parameter value in the data-generating mechanism $F_0$, which restricts the loss function that can be used but cannot guarantee the loss function to be unique. Furthermore, many of the arguments in support of the GB method given by \cite{bissiri2016general} depend on having a genuine loss function, but in pure parametric inference problems, it is often the case that we do not possess a genuine loss function $l(x,\theta)$ that represents the cost of discrepancy between the observed data $x$ and, in the language of decision theory, the decision/action $\theta$. Another problem, shared by both BEL and GB method, is that these systems of inference are incapable of generating predictive inference without further assumptions as they do not specify the distribution of the observables. Furthermore, in these pseudo-Bayesian systems the substitutions we make in place of the likelihood function depends on the target of inference, which brings up the question of whether any two ``probability" statements for two different parameters are jointly coherent, or do they violate the rules of probability as was the case with fiducial inference and confidence distributions \citep{lindley1958fiducial,cox2006principles}. Given the increased interest in and downsides of likelihood-free pseudo-Bayesian inference, we aim to find a fully Bayesian alternative to address these use cases. Our solution is to target functional parameters of the distribution for the observable while keeping the likelihood model fully nonparametric in order to avoid misspecification; see \cite{ghosh2003bayesian} or \cite{ghosal2017fundamentals} for general methods in Bayesian nonparametrics, and \cite{rubin1981bayesian} for the related method of Bayesian bootstrap. 
	
	Although nonparametric models have long been considered to be beneficial in avoiding misspecification \citep{hjort2010bayesian}, we typically observe incoherence between the subjective prior $p_\theta$ and the chosen nonparametric model. This arises due to the functional parameter being map from the space of distributions for the observable to some parameter space, therefore a nonparametric prior $Q$ over the space of distributions for the observable \textit{induces} a belief distribution for the functional parameter which seldom matches $p_\theta$.There is, however, a simple technique that will allow us to work with nonparametric models and at the same time ensure the required subjective marginal prior distribution for any functional parameter is reflected in the nonparametric prior. We refer to this method as $\theta$-augmentation. 
	
	The $\theta$-augmentation method, to be introduced shortly, allows us to fully specify a subjective prior for a functional parameter despite it not being a parameter in the nonparametric likelihood, thus ensuring conveyance of important prior information regarding the functional parameter and efficient inference. Through separation of parameter definition from likelihood parameterization, we may tackle with ease the problems that currently do not have well-accepted likelihoods, examples including causal inference and missing data problems. By adopting the $\theta$-augmentation method, Frequentist tools such as inverse probability weighting and matching can be easily ported to the Bayesian framework. Our method is also potentially of interest in areas of application that typically deal with low sample sizes, where Bayesian inference may be preferred over the Frequentist, based on its interpretation in finite-sample inference but where previous implementations of Bayesian models are either parametric (thus subject to misspecification) or nonparametric and lacking in precise control over the marginal subjective prior for the target parameter, see for example the meta-analysis of \cite{burr2005bayesian}
	
	In this paper, we will introduce the theory of $\theta$-augmentation, address techniques to sample from the posterior of a $\theta$-augmented model, discuss suitable and practical choices of nonparametric models on which to apply $\theta$-augmentation for consistent inference. We provide two simulation studies, the first of which is aimed at highlighting the performance of $\theta$-augmented model for Bayesian inference over competing likelihood-free pseudo-Bayesian methods under the simple setting of estimating the mean and variance, and the second of which is an application in inference with missing data, to demonstrate how one may achieve ``double robustness" that is hard to achieve with previous methods for Bayesian inference in missing data problems.
	
	\section[Theory of $\theta$-augmentation]{Theory of $\theta$-augmentation of models for Bayesian inference}\label{sec:theory}
	\subsection{Defining a $\theta$-augmented model}
	Let the target functional target parameter be denoted by $\theta$. Without loss of generality, let $M(\mathds{R})$ denote the space of all probability distributions over the real line, and $\Sigma$ a $\sigma$-algebra over this sample space. The underlying idea to the $\theta$-augmented model for semiparametric inference is to modify a convenient ``proposal'' probability model $\mathcal{P}_\Pi = (M(\mathds{R}), \Sigma, \Pi)$ according to the contours of the model space given by target parameter $\theta$. The proposal model $\mathcal{P}_\Pi$ is a model for the distribution for the observable random variable. The theory for $\theta$-augmentation presented here has two prerequisites, namely 1) infinite exchangeability of the random variables of which the observed data $(x_1,\ldots,x_n)$ are a finite realization, and 2) existence of a single dominating measure for all models in the prior model space \textit{almost surely}, see p.17 of \cite{ghosh2003bayesian}, which we shall refer to as the probability model being dominated.
	
	Suppose we have $\theta:M(\mathds{R})\rightarrow \mathds{R}^d$, $d\in \mathds{Z}^+$. Let there be a function $m:\mathds{R}^d \rightarrow \mathds{R}^+$ s.t. the composite function $m \circ \theta:M(\mathds{R})\rightarrow \mathds{R}^+$ is a $\Sigma$-measurable positive 
	function that is integrable, that is, $\int_{M(\mathds{R})} m(\theta(F_X)) \mathrm{d}\Pi(F_X)<\infty$. We define a new probability measure $\Pi^\star$, the $\theta$-augmented measure, to be,
	\begin{align}\Pi^\star(A) &= \frac{\int_{M(\mathds{R})} \mathds{1}[F_X \in A] m(\theta(F_X)) \mathrm{d}\Pi(F_X)}{\int_{M(\mathds{R})} m(\theta(F_X)) \mathrm{d}\Pi(F_X)}\label{eq:PiStar},
	\end{align}
	for any measurable event $A$ in $\Sigma$. This construction is conceptually similar to a slicing, or partitioning, of the model space based on $\theta(F_X)=t$ and, for every model in this sub-space, adjusting its probability content by a factor $m(t)$.
	
	We can verify that $\Pi^\star$ is a valid probability measure on $(M(\mathds{R}),\Sigma)$. Let the normalizing constant of $\Pi^\star$ be $
	Z_{\Pi^\star} = \int_{M(\mathds{R})} m(\theta(F_X)) \mathrm{d}\Pi(F_X)$. The measure of any event $A\in\Sigma$ under the unnormalized measure $\Pi^\star Z_{\Pi^\star}$ is equal to the expectation of the measurable non-negative function $m(\theta(F_X)) \mathds{1}[F_X \in A]$ taken under $\Pi$. The countable additivity of $\Pi^\star  Z_{\Pi^\star}$  can easily be verified. Therefore $\Pi^\star$ is a probability measure; it is a distribution over random measures and its definition depends on both $m$ and $\Pi$. 
	
	While we see Eqn. (\ref{eq:PiStar}) as a recipe for constructing semiparametric models from nonparametric ones via parameter augmentation, the corresponding change of measure can also be regarded as generalizing the usual approach to importance sampling, with $\Pi^\star$ playing the role of the target distribution and $\Pi$ the biasing distribution.  Typically, importance sampling is not performed with nonparametric distributions. Our construction of $\Pi^\star$ in Eqn. (\ref{eq:PiStar}) ensures that the Radon--Nikodym derivative of the target distribution $\Pi^\star$ with respect to $\Pi$ depends only on the functional parameter $\theta$; in this case that derivative is $(m\circ\theta)/Z_{\Pi^\star}$. 
	
	As a probability model for $F_X$, we define the $\theta$-augmented probability model, denoted as
	$
	\TA(m, \mathcal{P}_\Pi),
	$
	to be the probability model which inherits the sample space and $\sigma$-algebra of $\mathcal{P}_\Pi$, and is equipped with the measure $\Pi^\star$ as defined according to Eqn. (\ref{eq:PiStar}). We will refer to $m(\cdot)$ as the weighting function, and refer to $\mathcal{P}_\Pi$ as the proposal model, and $\Pi$ as the proposal measure. For simplicity, we will state $\Pi^\star$ as the probability measure of a $\TA(m, \mathcal{P}_\Pi)$ model, as often as needed, so that the parameters $m$ and $\mathcal{P}_\Pi$ associated with a $\Pi^\star$ are understood clearly. The domain of the weighting function $m$ will be indicated explicitly in its specification, to remove any ambiguity with regards to the construction of $\Pi^\star$. Appropriate choices of $\mathcal{P}_\Pi$ will be discussed in Section \ref{sec:proposal}. 
	
	One of our primary goals is to specify a $\theta$-augmented model such that the corresponding marginal prior distribution of $\theta$ is equal to our subjective prior. Whether the induced distribution is absolutely continuous with respect to the Lebesgue measure or discrete will depend on the particular $\mathcal{P}_\Pi$ and the functional of interest. The important point is that the induced marginal prior and our subjective prior should be dominated by the same measure. Below we give a prescription to achieve the desired marginal prior distribution for when $\mathcal{P}_\Pi$ induces an absolutely continuous distribution over $\theta$. The technique easily generalizes to the discrete case.
	
	Let $q^\Pi_\theta$ denoting the density function of $\theta$ induced by the proposal measure $\Pi$. Let the marginal subjective prior density be denoted by $p_\theta$. To achieve the correct marginal prior for a $\theta$-augmented model, we may choose $$m = p_\theta/q^\Pi_{\theta},\label{eq:m1}$$ where $p_\theta/q^\Pi_{\theta}$ denotes point-wise division of $p_\theta$ by $q^\Pi_{\theta}$. As a convention in this paper, if $f$ and $g$ are two functions with the same domain, we shall use $fg$ or $f\cdot g$ to denote the point-wise product of these functions, and $\frac{f}{g}$ or $f/g$ to denote point-wise quotient.
	
	We can verify that the correct marginal prior results. Denote by $F^{\Pi^\star}_\theta$ the distribution function of $\theta$ induced by the measure $\Pi^\star$ of the $\TA(m = p_\theta/q^\Pi_{\theta},\mathcal{P}_\Pi)$ model, we have that
	\begin{align}\label{eq:priorMarginal}
		F^{\Pi^\star}_\theta(t) &= \dfrac{ \int_{M(\mathds{R})} \mathds{1}[\theta(F_X) \leq t] \frac{p_\theta(\theta(F_X))}{q^\Pi_\theta(\theta(F_X))} \mathrm{d}\Pi(F_X)}{Z_{\Pi^\star}}\\
		& \propto \int_{-\infty}^{\infty} \mathds{1}[u \leq t] \frac{p_\theta(u)}{q^\Pi_\theta(u)} \cdot q^\Pi_\theta(u) \mathrm{d}u = \int_{-\infty}^t p_\theta(u) \mathrm{d}u,	 	\label{eq:priorMarginal-step2}
	\end{align}
	i.e. the density function corresponding to $F^{\Pi^\star}_\theta(t)$ is exactly $p_\theta$.
	The simplification from Eqn. (\ref{eq:priorMarginal}) to (\ref{eq:priorMarginal-step2}) is due to $ \mathds{1}[\theta(F_X) \leq t] p_\theta(\theta(F_X))/q^\Pi_\theta(\theta(F_X))$ depending only on the value of $\theta(F_X)$. 
	
	In the case that the distribution for $\theta$ induced by the proposal model is discrete, with the probability mass function (PMF) of our subjective prior be denoted by $\text{P}_\theta$ and the PMF of $\theta$ induced by $\mathcal{P}_\Pi$ denoted by $\text{Q}^\Pi_\theta$, choosing
	$m = \text{P}_\theta/\text{Q}^\Pi_\theta$
	as the weighting function achieves the required marginal distribution over $\theta$ under the TA model.
	
	\subsection{Posterior distribution under the $\theta$-augmented model}
	We first discuss the posterior distribution of $\theta\mid\tilde{x}_n$ under the TA model in the case that the distribution for $\theta$ induced by the proposal model is absolutely continuous with regards to the Lebesgue measure. Let $F_X \sim \TA(m = p_\theta/q^\Pi_{\theta}, \Pi)$ \textit{a priori}. Because we restrict ourselves to $\mathcal{P}_\Pi$ where every $F_X$ in the space of random measures is dominated by the same dominating measure \textit{almost surely}, Bayes rule applies. Let $q^\Pi_{\theta\mid\tilde{x}_n}$ represent the density function of $\theta\mid\tilde{x}_n$ induced by $\Pi$. We obtain $F^{\Pi^\star}_{\theta\mid\tilde{x}_n}$, the distribution function of $\theta\mid\tilde{x}_n$ induced by $\Pi^\star$, as
	\begin{align}
		F^{\Pi^\star}_{\theta\mid\tilde{x}_n}(t\mid\tilde{x}_n) &\propto  \int_{M(\mathds{R})}   \mathds{1}[\theta(F_X) \leq t] \cdot   \prod_{i=1}^n F_X(x_i)\mathrm{d}\Pi^\star(F_X)\label{eq:post_Bayes_theta}\\
		&\propto  \int_{M(\mathds{R})}   \mathds{1}[\theta(F_X) \leq t] \cdot  \frac{p_\theta(\theta(F_X))}{q^\Pi_\theta(\theta(F_X))}\cdot  \prod_{i=1}^n F_X(x_i)\mathrm{d}\Pi(F_X)\\
		& \propto \int_{M(\mathds{R})}   \mathds{1}[\theta(F_X) \leq t] \cdot  \frac{p_\theta(\theta(F_X))}{q^\Pi_\theta(\theta(F_X))}\cdot \mathrm{d}\Pi(F_X\mid\tilde{x}_n)\\
		&= \int_{-\infty}^t \frac{q^\Pi_{\theta\mid\tilde{x}_n}(u\mid\tilde{x}_n)}{q^\Pi_\theta(u)} \cdot p_\theta(u) \mathrm{d}u.\label{eq:post_Bayes_simplified}
	\end{align}
	Bayes rule is important here in order to deduce the form of the posterior distribution of $\theta$ under $\Pi^\star$. Had it not applied, the form of $F^{\Pi^\star}_{\theta\mid\tilde{x}_n}$ would have to be found through other means similar to other non-dominated models in Bayesian nonparametrics, and the weighting function $p_\theta/q^\Pi_\theta$ may not have appeared in $F^{\Pi^\star}_{\theta\mid\tilde{x}_n}$ as elegantly as it did in Eqn. (\ref{eq:post_Bayes_simplified}). 
	
	There are several observations to be made based on Eqn. (\ref{eq:post_Bayes_theta})-(\ref{eq:post_Bayes_simplified}). Firstly, let $q^{\Pi^\star}_{\theta\mid\tilde{x}_n}$ denote the target posterior density function corresponding to  $F^{\Pi^\star}_{\theta\mid\tilde{x}_n}$, we have that $$
	q^{\Pi^\star}_{\theta\mid\tilde{x}_n} \propto \left(q^\Pi_{\theta\mid\tilde{x}_n}/q^\Pi_\theta\right) p_\theta.
	$$
	This expression shows clearly that, after seeing the data, our marginal prior distribution $p_\theta$ is modified by $q^\Pi_{\theta\mid\tilde{x}_n}/q^\Pi_\theta,$ which we term the \textit{effective likelihood}. Secondly, the posterior distribution for $F_X$ is again distributed according to a $\theta$-augmented model, but now the proposal model is equipped with measure $\Pi(\cdot\mid\tilde{x}_n).$

	In the case that the distribution for $\theta$ induced by the proposal model is discrete, and assuming $F_X \sim \TA(m = \text{P}_\theta/\text{Q}^\Pi_{\theta}, \mathcal{P}_\Pi)$ \textit{a priori}, we have that 
	\begin{equation*}
		\Pi^\star(\theta=t\mid\tilde{x}_n) \propto \frac{\text{P}_\theta(t)}{\text{Q}^\Pi_\theta(t)}\text{Q}^\Pi_{\theta\mid\tilde{x}_n}(t\mid\tilde{x}_n),
	\end{equation*}
	for $t$ in the support of $\text{Q}^\Pi_\theta$.
	
	Parametric consistency of posterior inference under the $\theta$-augmented model obtains if the $\mathcal{P}_\Pi$ exhibits parametric consistency and the weighting function $m$ is bounded. The boundedness condition is typically satisfied as we require that $m$ be an integrable function under the measure $\Pi$, otherwise may be achieved by a truncation of the parameter space. The proof is straightforward and hence given in Appendix \ref{sec:asymptotics}, which also provides additional details on the parametric consistency of weakly consistent nonparametric models, for functional parameters defined via estimating equations.

	\subsection{Multivariate $\theta$ and marginal inference}\label{sec:multivariate}
	The method of $\theta$-augmentation is not limited to $\theta\in\mathds{R}^1$. In the case that the target parameter is $\theta = (\theta_1,\ldots, \theta_d) \in \mathds{R}^d$. As a simple extension to Eqn. (\ref{eq:priorMarginal}), a $\theta$-augmented model with the weighting function
	\begin{equation}
		m(\theta_1,\ldots, \theta_d ) = \frac{p_\theta(\theta_1 ,\ldots,\theta_d )}{q^\Pi_\theta(\theta_1 ,\ldots,\theta_d )},\label{eq:joint_inference_inducedprior}
	\end{equation}
	will lead to the prior marginal density of $\theta$ being exactly $p_\theta(\theta_1,\ldots,\theta_d)$ as required.

	We may also flexibly specify a subjective prior marginally for a subset of the parameter vector, if we lack well-formed knowledge for the whole parameter vector.  Suppose we partition the $d$-dimensional parameter vector $\theta$ as $$\theta = (T_1, T_2), \quad T_1\in\mathds{R}^{d_1},\quad  T_2 \in \mathds{R}^{d_2}, \quad d_1+d_2=d,$$ and we have a well-informed subjective prior for $T_1$ only. We may assert this partial information, i.e. requiring that we match marginal prior $p_{T_1}$ in our model, by letting
	\begin{align}
		m(\theta)= m(T_1,T_2) &= \frac{p_{T_1}(T_1)\times q^\Pi_{T_2\mid T_1}(T_2)}{q^\Pi_{T_1,T_2}(T_1, T_2 )}=\frac{p_{T_1}(T_1 )}{q^\Pi_{T_1}(T_1 )},\label{eq:joint_inference_partialspecification}
	\end{align}
	where $q^\Pi_{T_2\mid T_1}$ is the conditional distribution of the remaining parameters given $T_1$, induced by $\Pi$. Then, marginal prior of $T_1$ under the corresponding $\theta$-augmented model will be exactly $p_{T_1}$ as required, but for the parameters in $T_2$, marginal prior under this $\theta$-augmented model will be
	\begin{align*}
		\Pi^\star(T_2 \leq t_2) &= \int \int \mathds{1}[T_2 \leq t_2] \frac{p_{T_1}(T_1)}{q^\Pi_{T_1}(T_1)} q^\Pi_{T_1,T_2}(T_1,T_2)\mathrm{d}T_1 \mathrm{d}T_2\\
		&=\int_{-\infty}^{t_2} \left[\int_{-\infty}^\infty  q^\Pi_{T_2\mid T_1}(u\mid T_1) p_{T_1}(T_1) \mathrm{d}T_1\right] \mathrm{d}u.
	\end{align*}
	This shows that a construction of $\theta$-augmented model through Eqn. (\ref{eq:joint_inference_partialspecification}) automatically adjusts the marginal prior (and posterior) of untargeted dimensions of $\theta$ coherently based on the laws of probability. We see that prior knowledge regarding $T_1$ will be informative of $T_2$ if $q^\Pi_{T_2\mid T_1}$ depends on $T_1$ under the proposal measure $\Pi$. The above method for inferring multivariate $\theta$ while supplying a prior for a subset of $\theta$ is a coherent system for joint inference on the whole $\theta$ vector or marginally on any subset.


	\section{Sampling from the target posterior}\label{sec:algo_MCMC}
	Suppose that $\mathcal{P}_\Pi$ induces a distribution for $\theta$ that is absolutely continuous, and we conduct Bayesian inference based on the model
	\begin{align*}
		X \sim F_X, \quad F_X \sim \TA(m=p_\theta/q^\Pi_\theta, \mathcal{P}_\Pi).
	\end{align*}
	The Bayesian posterior for $\theta$ has the density
	$	q^{\Pi^\star}_{\theta\mid\tilde{x}_n} \propto \left(p_\theta/q^\Pi_\theta\right) q^\Pi_{\theta\mid\tilde{x}_n}
	$.
	In the case that the proposal model induces a distribution for $\theta$ that is discrete, Bayesian inference via the $\theta$-augmented model will have a very similar form as the above.  We therefore discuss sampling based on the continuous version of the posterior distribution for $\theta$, which easily generalizes to the discrete version. 
	
	
	When posterior inference on the target $\theta$ is required, the most efficient way to obtain samples from the marginal posterior density $q^{\Pi^\star}_{\theta\mid\tilde{x}_n}$ is direct sampling either via the inverse cumulative distribution function method, or grid approximation. Sampling via Markov chain Monte Carlo (MCMC) is also possible, and can be particularly advantageous when dealing with joint inference for multivariate $\theta$.
	
	A simple Metropolis--Hastings (MH) algorithm proposes moves in the space of $F_X$ based on the proposal measure \textit{a posteriori}, $\Pi(\cdot\mid\tilde{x}_n)$. In the case that independent sampling of $F_X$ from $\Pi(\cdot\mid\tilde{x}_n)$ is possible, the acceptance ratio of a move from $F_X$ to $F_X'$ is
	\begin{align}
		A(F_X', F_X) 
		&=\min\left(1, \frac{p_\theta(\theta(F_X'))/q^\Pi_\theta(\theta(F_X'))}{p_\theta(\theta(F_X))/q^\Pi_\theta(\theta(F_X))}\right). \label{eq:MH-acceptanceRatio}
	\end{align}
	In the case that Gibbs sampling, a special case of MH, from $\Pi(\cdot\mid\tilde{x}_n)$ is possible, we have that the acceptance ratio for each \textit{conditional} move is also according to Eqn. (\ref{eq:MH-acceptanceRatio}). Marginal inference is obtained by transformation of sampled $F_X$ to $\theta$. These two special cases of the MH algorithm are sufficient to address sampling from the posterior $\theta$-augmented model under the specific choices of $\mathcal{P}_\Pi$ we outline in Section \ref{sec:proposal}.
	
	Other MH proposal distributions are possible, but the advantage of using exactly $\Pi(\cdot\mid\tilde{x}_n)$ as the MH proposal is that the acceptance ratio involves only values of $\theta$ thus one need not keep track of the full $F_X$ when running the Markov chain. To satisfy the integrability requirement of the weighting function, we will most often choose a model $\mathcal{P}_\Pi$ that induces a $q^\Pi_\theta$ that is wider than our subjective prior $p_\theta$. As a result, the posterior $q^{\Pi^\star}_{\theta\mid\tilde{x}_n}$ will be less spread out than  $q^{\Pi}_{\theta\mid\tilde{x}_n}$, and the default choice of $\Pi(\cdot\mid\tilde{x}_n)$ as the proposal distribution for the MH algorithm is appropriate. The algorithm will be efficient if $q^{\Pi^\star}_{\theta\mid\tilde{x}_n}$ is similar to $q^{\Pi}_{\theta\mid\tilde{x}_n}$. 
	
	If closed form expressions for $q^\Pi_\theta$ and $q^{\Pi}_{\theta\mid\tilde{x}_n}$ are unavailable, a practical solution is to substitute with estimated versions obtained via common density estimation packages, e.g. the R library \textbf{ks} \citep{ks} for density estimation up to six dimensions. Prior to performing direct sampling, we require estimates of both  $q^\Pi_\theta$ and $q^{\Pi}_{\theta\mid\tilde{x}_n}$, whereas for MCMC sampling, we require only an estimate of $q^\Pi_\theta$. We found that, if we choose $\mathcal{P}_\Pi$ with the requirement that the induced $q^{\Pi}_{\theta\mid\tilde{x}_n}$ has essentially no weight in the tail regions of $q^\Pi_\theta$, then the use of estimated $\hat{q}^\Pi_\theta$ and $\hat{q}^{\Pi}_{\theta\mid\tilde{x}_n}$ in our proposed sampling schemes produces good samples for approximating the required target distributions.
	
	\section{Nonparametric proposals for the $\theta$-augmented model}\label{sec:proposal}
	As seen in Appendix \ref{sec:asymptotics}, one way to guarantee the parametric consistency of the $\theta$-augmented model is to ensure that the proposal model, $\mathcal{P}_\Pi$, is at least weakly consistent for the data-generating distribution. As such we may wish to avoid nonparametric Bayesian models based on conditional likelihoods to prevent misspecification.  We find the general class of infinite kernel mixture models to satisfy our needs, since the posterior distribution is known to converge to the joint data-generating distribution weakly \citep{ghosh2003bayesian} with minimal conditions. We also note that, as a prerequisite in the present theory, suitable nonparametric proposal models must necessarily be dominated, which means that every random distribution in the support of the proposal model is dominated by the same measure.
	
	The Dirichlet process mixture (DPM) of continuous kernels is a special case of the infinite kernel mixture model that is dominated and well-studied in terms of algorithms and asymptotics \citep{ghosal1999posterior, wu2010l1}. In the case that the observable is a vector with both discrete and continuous parts, there exists a DPM of general linear models \citep{hannah2011dirichlet} for estimation of the joint density/mass function, which is weakly consistent for the data-generating distribution under certain assumptions. With a DPM as the proposal model, posterior sampling from the target $\theta$-augmented model can be performed as recommended in Section \ref{sec:algo_MCMC} based on blocked Gibbs sampling from the DPM \citep{ishwaran2001gibbs}. If we consider the extra computation required on top of standard Gibbs sampling from $\mathcal{P}_\Pi$ in order to obtain inference under the $\theta$-augmented model, this additional computation is low in complexity for several important use cases, including estimation of the coefficients of linear regression. But in general we may be presented with a significant amount of additional computational burden depending on the form of $\theta$. As an example, a functional defined as the solution to $\int g(x, t) \mathrm{d}F_X(x) = 0$, where $g$ is nonlinear, will likely not have a closed-form expression, and therefore we may need to solve this estimating equation through numerical means every time we find ourselves in need to map a sampled random mixture distribution to $\theta$.

	In order to infer with ease any general functional parameter based on any type of observable, we consider an alternative type of infinite kernel mixture model--- the Dirichlet process (DP), which is known to be weakly consistent for the data-generating distribution \citep{ghosal2017fundamentals}. The typical Dirichlet process supported on the real line is not dominated, thus violating a prerequisite of the $\theta$-augmented model. A simple adjustment is to instead consider the data as recorded to finite precision which may be modelled with a Dirichlet process with a discrete base distribution, which is dominated.  The practical implications of rounding on finite sample inference is minimal if the precision is high, especially considering that most computers perform finite precision arithmetic unless otherwise specified. We find the Dirichlet process with a discrete base distribution particularly useful as it can be applied to model observable vectors with both discrete and continuous parts. The discreteness of random measures in the sample space of the Dirichlet process makes transformation of random $F_X$ to $\theta$ straightforward, which, combined with the availability of independent sampling for DP, results in efficient posterior sampling from the $\theta$-augmented model. 
	
	
	\section{Simulation studies}\label{sec:sims}
	\subsection{Joint estimation of the mean and variance parameters}\label{sec:compare-meanvar}
	Bayesian inference via the $\theta$-augmented model, or $\theta$-augmented Bayesian inference for short, is asymptotically well justified. We performed a simulation study to highlight the strengths of $\theta$-augmented Bayesian inference in small sample inference against three competitors, namely the Bayesian bootstrap \citep{rubin1981bayesian}, Bayesian empirical likelihood \citep{lazar03}, and the general Bayes method \citep{bissiri2016general}. 
	
	We considered the problem of mean and variance estimation from exchangeable random observables $X_{i}$, $i=1,\ldots,n$. The target parameter is therefore $\theta=(\mu,\sigma^2)$. In the case of general Bayes method we identified the joint parameter via m-estimation with tuning weights chosen based on asymptotic tuning; see Appendix \ref{sec:SupplmentDetails_meanvar}. For $\theta$-augmented Bayesian inference, we used a $\theta$-augmented model with $m= p_{\mu,\sigma^2}/q^\Pi_{\mu,\sigma^2}$, and $\mathcal{P}_\Pi$ being a Dirichlet process with the precision of 0.5 and a base distribution that is a discretized version of the $\text{Normal}(\mu_0=0,\sigma_0^2=100)$ distribution. Posterior sampling for $\theta$-augmented Bayesian inference is performed using the MH algorithm in Section \ref{sec:algo_MCMC}. 
	
	Details regarding the competing methods, definition of evaluation metrics, model specifications, implementations, and various sampling procedures undertaken are provided in Appendix \ref{sec:SupplmentDetails_meanvar}. Here we briefly mention that the data is generated according to a skewed distribution, where for all $i \in \lbrace 1 ,\ldots,n\rbrace$,
	\begin{align*}
		X_i = -6 Z_i + T_i, \quad
		T_i \sim \text{Normal}(\mu= 5, \sigma^2= 4 ),\quad
		Z_i \sim \text{Bernoulli}(0.25),
	\end{align*} and that we use the same joint prior, $$	1/\sigma^2 \sim \text{Gamma}(\alpha= 6.623, \beta= 60.442), \quad
	\mu\mid \sigma^2 \sim \text{Normal}(\mu_0=3.5, \sigma_0^2=\sigma^2 ),
	$$
	for all methods requiring a subjective prior distribution; this prior was specified such that the marginal prior mean of variance parameter matching the variance of data-generating distribution, and the marginal prior mean of mean parameter matching the mean of data-generating distribution.

	\begin{table*}
			\centering
		\caption{Results for estimating $\mu$ and $\sigma^2$ jointly, Section \ref{sec:compare-meanvar}. Performance metrics were estimated based on 300 datasets at each sample size $n$, based on joint highest density credibility regions. The largest Monte Carlo standard errors for results in columns 3 and 4 are less than 0.03 and 1 respectively}
		\label{tab:compare-meanvar_2d}
		{
			\begin{tabular}{@{}cccccc@{}}
				\hline
				& & Coverage prob. & Average size \\
				Sample size & Method& of 95\% joint CR& of 95\% joint CR \\
				\hline
				{$n= 20$} &TAB & 0.98 & 30  \\
				&BB & 0.80  & 30 \\
				&BEL &  0.84&  26 \\
				&GB & 0.78 & 26  \\
				[6pt]
				{$n= 50$}  &TAB & 0.96 & 15 \\
				&BB & 0.88  & 14   \\
				&BEL &  0.89& 13  \\
				&GB & 0.87 & 13 \\
				\hline
		\end{tabular}}
	\end{table*}

	\begin{table*}
		\centering
		\caption{Results for marginal inference of $\mu$, Section \ref{sec:compare-meanvar}. Performance metrics were estimated based on 300 datasets at each sample size $n$. The largest Monte Carlo standard errors for results in columns 3 to 6 are less than 0.02, 0.03, 0.05 and 0.05 respectively}
		\label{tab:compare-meanvar_mean}{
			\begin{tabular}{@{}cccccc@{}}
				\hline
				& & Coverage prob. & Average size & Absolute & Average\\
				Sample size & Method& of 95\% CR& of 95\% CR & bias & quadratic risk \\
				\hline
				{$n= 20$} &TAB & 0.95& 2.61 & 0.05 &0.92\\
				&BB &  0.93 &2.70  &  0.10&   1.03  \\		
				&BEL & 0.95& 2.59 &  0.08& 0.91\\		
				&GB &0.93& 2.63& 0.06& 0.93 \\
				[6pt]
				{$n= 50$}  &TAB & 0.94 &1.73 &0.04&0.40 \\	
				&BB & 0.92 &1.76  &0.01   &  0.42 \\		
				&BEL &  0.94 &  1.79 &0.02&  0.42 \\		
				&GB &0.94 &1.77 &0.05&0.41\\
				\hline		
		\end{tabular}}
		
	\end{table*}
	
	\begin{table*}
		\centering
		\caption{Results for marginal inference of $\sigma^2$, Section \ref{sec:compare-meanvar}. Performance metrics were estimated based on 300 datasets at each sample size $n$. The largest Monte Carlo standard errors for results in columns 3 to 6 are less than 0.03, 0.03, 0.2 and 1.3 respectively}
		\label{tab:compare-meanvar_var}{
			\begin{tabular}{@{}cccccc@{}}
				\hline
				& & Coverage prob. & Average size & Absolute & Average\\
				Sample size & Method& of 95\% CR& of 95\% CR & bias & quadratic risk \\
				\hline
				{$n= 20$} &TAB & 1.00 &9.6& 0.6&11\\		
				&BB & 0.83  &  9.6  & 1.0  &   17  \\		
				&BEL & 0.88  & 8.3 &0.9 & 10 \\		
				&GB &0.82 &8.4& 1.6 &11 \\
				[6pt]
				{$n= 50$}  &TAB & 0.98 & 7.2 &0.4 & 6 \\		
				&BB & 0.88  &7.1 &0.6 &  8    \\		
				&BEL &  0.91 &6.4  &0.5&6 \\		
				&GB &0.86&6.8 &1.1& 6\\
				\hline		
		\end{tabular}}	
		\end{table*}

		The results of our simulation are summarized in Table \ref{tab:compare-meanvar_2d} for joint inference and Table \ref{tab:compare-meanvar_mean} and Table \ref{tab:compare-meanvar_var} for marginal inference. The results for the Bayesian empirical likelihood method were obtained based on approximations to posterior distributions with a low number of Monte Carlo samples, due to high computational demand, and is therefore to be interpreted with a larger margin of error.
		
		
		
		In joint inference, only $\theta$-augmented Bayesian inference achieved a coverage level close to nominal at the sample sizes tested. At any given sample size, the size of 95\% credibility region of $\theta$-augmented Bayesian inference was similar to that of competing methods while achieving much better coverage. For marginal inference on the variance parameter, out of all methods investigated, the $\theta$-augmented Bayesian inference had the smallest bias and the closest estimated coverage probability to nominal level, at any given sample size. Since variance is a function of the first and second moments, poor performance of general Bayes posterior inference was likely due to the skewness of the distribution of $X^2$ (see Figure \ref{fig:data-gen-comparisom-meanvar} of Appendix \ref{sec:SupplmentDetails_meanvar}) conflicting with the assumptions of the general Bayes method. Similarly, as posteriors for Bayesian bootstrap and Bayesian empirical likelihood are related to the empirical distribution, the empirical distribution of $X^2$ was likely unrepresentative of the population. Interestingly, the $\theta$-augmented Bayesian inference may have accounted for this with the incorporation of the weighting function $p_{\mu,\sigma^2}/q^\Pi_{\mu,\sigma^2}$; the induced proposal prior $q^\Pi_{\mu,\sigma^2}$ captures the skewness of the variance parameter \textit{a priori} due to sampled $F_X$ from the prior proposal model assigning large weights on a small number of support points, which mimics taking a small sample from the population.
		
		In marginal inference of the mean, all methods performed relatively well. The results presented in Table \ref{tab:compare-meanvar_mean} did not highlight one method above others as being uniformly superior.

		\subsection{Estimating the mean with data missing at random}\label{sec:MAR}
		We consider the problem of estimating the mean of a random variable $Y$ when some observations are missing at random, while an auxiliary variable $X$ is observed always. Let $C$ be the indicator for whether variable $Y$ is observed. When $Y$ is said to be missing at random, it means that $C \perp Y\mid X$, or equivalently, that ${Pr}(C\mid X,Y) = {Pr}(C\mid X)$. The observed data is the collection $(c_iy_i, c_i, x_i)$, $i = 1, \ldots, n$. The problem was studied by \cite{ray2020semiparametric} via a semiparametric conditional likelihood based approach which was not ``doubly robust".
		
		We can obtain doubly robust semiparametric Bayesian inference by using the $\theta$-augmented model if we define the functional parameter appropriately.  The key here is to borrow directly from the Frequentist literature a consistent estimator that can be viewed as a functional of the empirical distribution. Let $F_n$ denote the empirical distribution of a sample of size $n$, and ${E}_{(F_n)}$ denote an expectation taken with respect to $F_n$. We define an augmented inverse propensity weighted (AIPW) estimator \citep{robins1994estimation} as	
		\begin{equation}
			\mu_{\text{AIPW}}(F_n) = {E}_{(F_n)}\left[ \hat{\zeta}(X; \hat{\beta_0}(F_n), \hat{\beta_1}(F_n))+ C\left\lbrace \dfrac{Y - \hat{\zeta}(X; \hat{\beta_0}(F_n), \hat{\beta_1}(F_n))}{\hat{p}(X; \hat{\psi}_0(F_n), \hat{\psi}_1(F_n))} \right\rbrace\right], \label{eq:AIPW}
		\end{equation}
		where $\hat{p}(x;\psi_0, \psi_1) = \left\lbrace1 + e^{-(\psi_0+\psi_1 x)}\right\rbrace^{-1}$, and $\hat{\zeta}(x;\beta_0,\beta_1) = \beta_0 + \beta_1 x$, and 
		\begin{equation}
			\left(\hat{\beta}_0(F_n), \hat{\beta}_1(F_n)\right) = \left\lbrace (b_0,b_1)\in {R}^2: {E}_{(F_n)}\left[(1\quad X)^\top \left\lbrace Y- (b_0+b_1 X)\right\rbrace\mid C=1\right] = 0 \right\rbrace, \label{eq:lin_reg_AIPW}
		\end{equation}
		\begin{equation}
			\left(\hat{\psi}_0(F_n), \hat{\psi}_1(F_n)\right) = \left\lbrace (s_0,s_1)\in {R}^2:  {E}_{(F_n)}\left[(1 \quad X)^\top\left(C - \dfrac{1}{ 1  + e^{ - (s_0 + s_1 X)}} \right)\right] = 0 \right\rbrace. \label{eq:log_reg_AIPW}
		\end{equation}
		The Frequentist estimator $\mu_{\text{AIPW}}(F_n)$ will be consistent if either $Y\mid X$ followed a linear model or $C\mid X$ follow a logistic regression model in the true data-generating distribution. Eqn. (\ref{eq:AIPW}) - (\ref{eq:log_reg_AIPW}) also provide the form of the functional target parameter, $\mu_{\text{AIPW}}(\cdot)$, which can take any distribution for the observable as its argument. The notation ${E}_{(\cdot)}$, which appears throughout Eqn. (\ref{eq:AIPW}) - (\ref{eq:log_reg_AIPW}), is used here to emphasize a change in the measure of an expected value when we pass in a different distribution as the argument to $\mu_{\text{AIPW}}(\cdot)$. Under the Bayesian paradigm, a probability distribution over the space of distributions for the observable induces a probability distribution for $\mu_{\text{AIPW}}$. 
		
		As for an appropriate Bayesian prior, we note that a Dirichlet process with discrete base distribution is weakly consistent for the data-generating distribution \citep{ghosh2003bayesian}. Using this as the proposal model of a $\theta$-augmented model will result in consistent posterior parametric inference if either $Y\mid X$ followed a linear model or $C\mid X$ follow a logistic regression model in the data-generating distribution, according the results in Section \ref{sec:asymptotics} and the Appendix. Here we assume that real-world measurements are of finite precision, so that the model space is dominated as required.
		
		The remainder of this section provides some insight into the small sample Frequentist performance of the Bayesian posterior of $\mu_{\text{AIPW}}$ under the $\theta$-augmented model, and demonstrates double robustness. In a simulation study, we compared the performance of the $\theta$-augmented model for inferring $\mu_{\text{AIPW}}$ under four data-generating mechanisms all with the unconditional mean of $Y$ of 6. Model \ref{model:MAR-LogisticRegression} has conditional means matching $\hat{p}$ and $\hat{\zeta}$, Model \ref{model:MAR-LogisticRegression-wronglinear} has a conditional mean of $Y\mid X$ that is misspecified, Model \ref{model:MAR-notLogisticRegression-linear} has a conditional probability $Pr(C=1\mid X)$ that is misspecified, and, finally, Model \ref{model:MAR-notLogisticRegression-nonlinear} has both $Pr(C=1\mid X)$ and conditional mean of $Y\mid X $ both misspecified. The details for these models are found in Appendix \ref{append:Ex2}.
		
		Bayesian inference was performed with the following model:
		\begin{align*}
			(X,C,CY) \sim F_{X,C,CY}, \quad F_{X,C,CY} \sim \TA(m=p_{\mu_{\text{AIPW}}}/q^\Pi_{\mu_{\text{AIPW}}}, \mathcal{P}_\Pi = \text{Model \ref{model:tabProp-MAR}}).
		\end{align*}
		The marginal prior for $\mu_{\text{AIPW}}$ was assumed to follow a $\text{Normal}(\mu=6,\sigma^2=50)$ distribution.  The proposal model was a Dirichlet process with a discrete base measure, with specifications given in Model \ref{model:tabProp-MAR} in Appendix \ref{append:Ex2}. This proposal model induced a prior $q^\Pi_\theta$ that was sufficiently diffuse compared to the posterior, thus we do not observe any issues from the use of estimated density $\hat{q}^\Pi_\theta$ in the Metropolis--Hastings sampling algorithm. The performance metrics we used in this section are the same as those of Section \ref{sec:compare-meanvar}. Results for $\theta$-augmented Bayesian inference over repeated datasets are summarized in Table \ref{tab:compare-MAR-BB-correctModel}. Based on the results, we see that the $\theta$-augmented model generated posterior inference with good finite sample properties, and a convergence towards the true mean, except in the case of Model \ref{model:MAR-notLogisticRegression-nonlinear} where the Frequentist estimator $\mu_{\text{AIPW}}({F}_n)$ would not be consistent either.

		\begin{table*}
			\centering
			\caption{Evaluation of the $\theta$-augmented Bayesian inference of the mean of $Y$ with data missing at random with the $\mu_{\text{AIPW}}$ functional. Datasets were generated according four mechanisms, Model \ref{model:MAR-LogisticRegression}- \ref{model:MAR-notLogisticRegression-nonlinear}, which are detailed in Appendix \ref{append:Ex2}. Performance metrics were estimated based on 500 or more datasets at each sample size $n$. The largest Monte Carlo standard errors for the results in columns 3 to 6 are less than 0.02, 0.2, 0.04 and 0.5 respectively}
			\label{tab:compare-MAR-BB-correctModel}
			{	
				\begin{tabular}{cccccc}
					\hline
					Model &Sample size & CP of 95\% CR & Avg. size of 95\% CR & Abs. bias & Avg. quadratic risk\\
					\hline
					{\ref{model:MAR-LogisticRegression}}& $20$ &  0.93        &    5.3    &   0.12  & 3.9   \\
					&$50$ & 0.94       & 3.3      & $<0.01$  &  1.5     \\
					&$ 100$ & 0.93 & 2.3 & $<0.01$ &0.7 \\
					&$ 500$ &0.95 & 1.0& $<0.01$ &  0.1\\
					[6pt]
					{\ref{model:MAR-LogisticRegression-wronglinear}}&$ 20$ &  0.89   &  8.5        & 0.61 &  11.2    \\
					&$ 50$ &  0.92   & 5.9  & 0.20 & 5.6 \\
					&$ 100$ & 0.90 & 4.5 & 0.11 &  3.9 \\
					&$ 500$ & 0.91 & 2.3 & $<0.01$ &  0.8\\
					[6pt]
					{\ref{model:MAR-notLogisticRegression-linear}}&$ 20$ &  0.93    &   5.5      &   0.09   &  4.4    \\
					&$ 50$ &  0.94     &  3.4  & 0.01  & 1.8   \\
					&$ 100$ & 0.94 & 2.4 & 0.02 & 0.8 \\
					&$ 500$ & 0.97 & 1.1 & 0.01  & 0.2\\
					[6pt]
					{\ref{model:MAR-notLogisticRegression-nonlinear}}&$ 20$ &  0.89    &  9.4        & 1.09  &  13.7   \\
					&$ 50$ &   0.83    & 6.6   & 0.72  &  8.6   \\
					&$ 100$ & 0.78 & 5.2& 0.68  &  5.5 \\
					&$ 500$ & 0.76 & 3.0 & 0.40 &  2.1 \\
					\hline
			\end{tabular}}
			
	\end{table*}

	\section{Discussion and conclusion}

	
	
	In this paper, we developed the method of Bayesian semiparametric inference via a $\theta$-augmented model for inferring functional parameters. The $\theta$-augmentation method presented in Section \ref{sec:theory} provides a way to adjust any dominated proposal model via a simple weighting ratio to achieve the desired marginal prior distribution for a particular functional target parameter without requiring it being part of the likelihood specification. Construction of a $\theta$-augmented model from some proposal model is achieved via the usual importance sampling formula. Inference via a $\theta$-augmented model is easy to implement computationally according to the proposed algorithms. Section \ref{sec:multivariate} highlights the flexibility of the method in modelling partial information on a multivariate target parameter, which has benefits in real world situations where a subset of the parameter is more studied than others, perhaps in regression analysis with a number of covariates. We are also free to target a low dimensional parameter even when a large number of nuisance parameters are involved in defining the target, e.g. in estimation of skewness, requiring only marginal prior for valid statistical inference.
	
	Though nonparametric models already exist for functional parameter estimation, we typically do not have full control over the marginal prior for the functional parameter, which implies that the opportunity to incorporate useful information is being overlooked. The simulation study in Section \ref{sec:compare-meanvar} stands to remind us that addition of well-informed prior information for the target functional can lead to better finite-sample performance and more efficient inference, if we compare the performance of nonparametric BB with that of the semiparametric $\theta$-augmented model.

	The method of Bayesian inference via the $\theta$-augmented model may be used to translate many existing Frequentist tools into a Bayesian setting, provided that the Frequentist estimator is expressed as a functional of the empirical data distribution. This is an important advantage as Frequentist estimators need not be motivated by the likelihood function. Although we are accustomed to tackling Bayesian inference by firstly analyzing the structure of the sampling distribution/likelihood function, writing it down explicitly in terms of the target parameter, this approach can produce less than desirable posterior inference when the likelihood function is misspecified or unavailable. However, it could be that the problem has a simple and well accepted Frequentist solution. For example, the problem of confounding in treatment effects is typically addressed by inverse probability of treatment (IPT) weighting under a Frequentist framework. Application of the $\theta$-augmented model to this type of problem removes the need for correct structuring of the likelihood function. Further work is needed to understand and quantify any additional advantages with regards to performance of $\theta$-augmented Bayesian inference compared with likelihood-based Bayesian methods. Work remains to demonstrate $\theta$-augmented Bayesian inference with real datasets and under various other settings where an appropriate likelihood function may not exist. 
	
	Since the posterior under a $\theta$-augmented model is proportional to $q^{\Pi}_{\theta\mid\tilde{x}_n}\cdot p_\theta/q^\Pi_\theta$, when we employ the Dirichlet process model as the proposal model, we can tune the function $q^{\Pi}_{\theta\mid\tilde{x}_n}$ towards the Bayesian bootstrap by choosing DP precision to be close to zero. This may be a good strategy for specifying a proposal model for data analysis as the Bayesian bootstrap is known to be well-behaved. Though in implementations with the DP as the proposal model, discretization of the base distribution is required, it may be justified by the practical limitation that all real world measurements must be made with finite precision, and is likely not an issue in small sample inference. Of course, should one disagree with the discreteness in distributions sampled from the DP, they may choose the Dirichlet process mixture model to be the proposal model, which is equally valid theoretical but with higher computation requirement.
	
	While a great variety of functional parameters are suitable for Bayesian inference with the $\theta$-augmented model, there is currently a limit as to the type of feasible proposal models.  By requiring that the proposal model of a $\theta$-augmented model be necessarily dominated, we are limited in the class of nonparametric models that can be used to describe the observable random variable. Extension of the theory of $\theta$-augmented Bayesian inference may be interesting to see how this requirement may be relaxed to make more nonparametric methods available as the starting point of Bayesian semiparametric inference for functional parameters.
	
	The method may be challenging to implement when a high-dimensional joint prior distribution must be satisfied for a high-dimensional functional parameter. In most problems, the denominator of the weighting function, $q^\Pi_\theta$, of a $\theta$-augmented model has to be estimated. To this end, we may employ the \verb*|R| library \textbf{ks}, which provides functions for density estimation in up to six dimensions. Further research is needed to tackle this problem either from the algorithmic or theoretical front. Also relating to algorithms, further research on algorithmic efficiency of $\theta$-augmented Bayesian inference in large samples is helpful to determine the applicability of the $\theta$-augmented model to large datasets.
	
	Finally, choosing targets of inference that are functionals of the distribution for the observable provides a simple way to avoid model misspecification and counters the perception that the Bayesian paradigm is more limiting than the Frequentist simply because it always requires a likelihood. The simple technique of $\theta$-augmentation provides a mechanism to transfer one's subjective knowledge regarding a functional parameter, just as we are trained to do so for a likelihood parameter, to facilitate efficient Bayesian inference and a complete coherence of the Bayesian model while offering the opportunity to work with a completely nonparametric model space.


	\section*{Acknowledgement}
	The authors would like to thank the Natural Sciences and Engineering Research Council of Canada for their support.
	
	
	\vspace*{-10pt}
	
\appendix
	\section{Asymptotics}\label{sec:asymptotics}
	\subsection*{General consistency and distributional results}
	Consistency of a Bayesian procedure for a parameter typically refers to a type of convergence in probability. \cite{ghosal1997review} defines consistency for a Bayesian posterior $P(\cdot\mid\tilde{x}_n)$ for a parameter as, for every neighbourhood $U$ of $\theta_0$
	\begin{equation}
		\lim_{n\rightarrow\infty}  P( U \mid\tilde{x}_n) = 1  \quad a.s. \quad \mathds{P}_{F_0},\label{eq:BayesConsistencyDef}
	\end{equation}  
	where $F_0$ denotes the data-generating distribution, and $\theta_0 = \theta(F_0)$.
	
	It is easy to see that the consistency of a $\theta$-augmented model depends on the proposal model. Presume we conduct Bayesian semiparametric inference with the  $\theta$-augmented model $\TA(m= p_\theta/q^\Pi_\theta, \mathcal{P}_\Pi).$ Suppose that the proposal model $\mathcal{P}_\Pi$ equipped with measure $\Pi$ satisfies consistency per complement of Eqn. (\ref{eq:BayesConsistencyDef}), that is, for every neighbourhood $U$ of $\theta_0$,
	$
	\lim_{n\rightarrow\infty}  \Pi( \theta \in U^c \mid\tilde{x}_n) = 0  \quad a.s. \quad \mathds{P}_{F_0}.
	$
	If the particular weighting function $p_\theta/q^\Pi_\theta$ is bounded by a number $M$, then clearly the induced distribution of $\theta$ under the posterior $\Pi^\star(\cdot\mid\tilde{x}_n)$ is consistent for $\theta_0$ as
	\begin{align*}
		\lim_{n\rightarrow\infty}  \Pi^\star( \theta \in U^c \mid\tilde{x}_n) &=\lim_{n\rightarrow\infty} \int \mathds{1}\left[ \theta(F_X) \in U^c\right]\cdot p_\theta(\theta(F_X))/q^\Pi_{\theta} (\theta(F_X)) \mathrm{d}\Pi(F_X\mid\tilde{x}_n)\\
		&\leq \lim_{n\rightarrow\infty} M\int \mathds{1}\left[ \theta(F_X) \in U^c\right] \mathrm{d}\Pi(F_X\mid\tilde{x}_n) = 0
	\end{align*}
	We can ensure to satisfy $p_\theta/q^\Pi_\theta< M$ because, as we need to ensure integrability of the weighting function in a $\theta$-augmented model, in practice we will almost always choose a proposal measure where the induced $q^\Pi_\theta$ is more heavy-tailed than $p_\theta$, satisfying the boundedness condition.
	
	Given a general valid weighting function $m$, the form of TAB posterior, which is proportional to $m\cdot q^\Pi_{\theta|\tilde{x}_n}$, has clear parallels in standard Bayesian inference with $q^\Pi_{\theta|\tilde{x}_n}$ taking the place of the likelihood function; see p. 287 of \cite{bernardo1994bayesian} for a development of asymptotics in standard Bayesian inference.  Suppose that $q^\Pi_{\theta|\tilde{x}_n}$ has a unique maximum $t_{q,n}$, and a Hessian matrix at $t_{q,n}$ of $\Sigma_n^{-1}$. We may expand $\log q^\Pi_{\theta|\tilde{x}_n}$ about its maximum $t_{q,n}$ to obtain $\log q^\Pi_{\theta|\tilde{x}_n}(\theta)= \log q^\Pi_{\theta|\tilde{x}_n}(t_{q,n} ) -\frac{1}{2} (\theta- t_{q,n})^\top (\Sigma_n^{-1}) (\theta- t_{q,n}) + R_n,$ with $R_n$ being the remainder term. Suppose that $m(\theta)$ has a unique maximum $t_0$, and a Hessian matrix at $t_0$ of $H_0$. An expansion of $\log m $ about its maximum $t_0$ gives
	$
	\log m(\theta) = \log m(t_0) - \frac{1}{2}(\theta-t_0)^\top H_0 (\theta-t_0) + R_0,
	$  
	where $R_0$ is some remainder term. We assume regularity conditions that make $R_n$ and $R_0$ negligible for large $n$. Then, $q^{\Pi^\star}_{\theta|\tilde{x}_n}$ (up to proportionality constant) may be approximated by
	\begin{align*}
		\exp\left(-\frac{1}{2} (\theta- t_{q,n})^\top (\Sigma_n^{-1}) (\theta- t_{q,n})  - \frac{1}{2}(\theta-t_0)^\top H_0 (\theta-t_0) \right)
	\end{align*}  for large $n$, which is the kernel of a $\text{Normal}(t_n, H_n)$ distribution, where $H_n=H_0+\Sigma_n^{-1}$ and $ t_n = H_n^{-1}\left(H_0 t_0 + \Sigma_n^{-1} t_{q,n} \right).$
	If we have parametric consistency under the proposal model, then the curvature of $\log q^\Pi_{\theta|\tilde{x}_n}$ about its maximum as given by $\Sigma_{n}^{-1}$ must increase with sample size. For large enough $n$, $H_0$ will be negligible compared to $\Sigma_{n}^{-1}$, which warrants an approximation of $q^{\Pi^\star}_{\theta|\tilde{x}_n}$ based on $\text{Normal}(t_{q,n}, \Sigma_n^{-1})$, which is also an approximation for the posterior for $\theta$ under the proposal model.
	
	It is clear that asymptotic behaviour of inference based on the $\theta$-augmented model depends on that of the proposal model. In particular, we can deduce parametric consistency of nonparametric proposal models that are weakly consistent for the data-generating mechanism, for functionals defined via continuous estimating equations.
	
	\subsection*{Parametric consistency of $\theta$-augmented Bayesian posterior for parameters defined via estimating equations when the proposal model is weakly consistent}\label{append:asymptotics_weaklyconsistentPi}
	
	Conditions for asymptotic consistency of Bayesian nonparametric models can be found throughout standard references in Bayesian nonparametrics, e.g.  \cite{ghosal2017fundamentals} and \cite{ghosh2003bayesian}. In this section we identify the conditions for parametric consistency of Bayesian posteriors based on $\theta$-augmented models, for functional parameters defined via estimating equations. We assume weak consistency of the nonparametric proposal model. We assume that the data-generating distribution has a density $f_0$, and the nonparametric proposal probability model with the measure $\Pi$, is supported on random measures admitting density functions $f$. Generalization of the results to probability  models supported on random distribution functions is straight forward. 
	
	We have that, for functional parameters defined via an estimating equation $g(x,t)$, that is,
	$ \theta(f) = \left\lbrace t \in \Theta \quad s.t. \int g(x, t) f(x) \mathrm{d}x=0 \right\rbrace$,
	boundedness of $g(x,t)$, continuity of $g(x,t)$ in $x$ and $t$, integrability of $\int g(x,t) f(x) \mathrm{d}x$ for all $f$ in the support of $\Pi$, integrability of $\int g(x,t) f_0(x) \mathrm{d}x$, and weak consistency of the proposal model together imply parametric consistency under the proposal model.
	
	This stems from the definition of weak consistency (Definition 1 of \cite{ghosal1999posterior}). Given bounded continuous functions $\phi_i, i=1,\ldots,k$, for $\epsilon>0$, let 
	\begin{equation}
		U = \left\lbrace f \in \mathcal{F}: \left|\int \phi_i(x)f(\mathrm{d}x) - \int \phi_i(x)f_0(\mathrm{d}x)\right|< \epsilon, i=1,2,\ldots,k \right\rbrace \label{def:weakNeighbourhood}
	\end{equation} be a weak neighbourhood of $f_0$. A probability model with measure $\Pi$ is said to be weakly consistent for $f_0$ if with $P_{f_0^-}$ probability 1, 
	$
	\Pi(U\mid X_1,X_2,\ldots, X_n) \rightarrow 1
	$ for \textit{all} weak neighbourhoods $U$ of $f_0$. The boundedness condition for $\phi_i$'s that define a weak neighbourhood may not be satisfied by $g(x,t)$ when the domain of the function is not compact at any given $t$. In this case, a work around is to truncate $x$ to be within some bounds -- this is generally possible to do when the observables are natural/physical phenomena. It is therefore assumed that $g(x,t)$ at any given $t$ can be used to define weak neighbourhoods in the development of the subsequent proof.
	
	For estimating equations that lead to explicit solutions for the parameter, in the form of $\theta(f) = \int k(x) f(\mathrm{d}x)$ for some general function $k(x)$, parametric consistency is immediately apparent. For example, when
	$\theta(f) := \int x f(\mathrm{d}x)$, weak consistency of a model for $f_0$ implies, for all $\epsilon > 0$, $$\lim_{n\rightarrow\infty} \Pi\left(\left\lbrace f \in \mathcal{F}: \left|\int x f(\mathrm{d}x) - \int x f_0(\mathrm{d}x)\right|< \epsilon \right\rbrace \mid X_1,X_2,\ldots,X_n\right) = 1,$$
	due to $\left\lbrace f \in \mathcal{F}: \left|\int x f(\mathrm{d}x) - \int x f_0(\mathrm{d}x)\right|< \epsilon \right\rbrace$ being a weak neighbourhood per definition given in Eqn. (\ref{def:weakNeighbourhood}). Similar arguments apply to show parametric consistency for estimating the $k$-th moment, variance/covariance, coefficients for linear regression or, in general, simple functions of moments when Slutsky's theorem applies.
	
	Now consider the case that the parameter $\theta$ is defined as the solution to some estimating equation. Without loss of generality assume $\theta\in \mathds{R}$. Let
	$	h_0(t) = \int g(x,t) f_0(x) \mathrm{d}x$ such that $h_0(\theta_0)= 0$, i.e. the estimating function identifies the true parameter $\theta_0$ under the true data generating mechanism $f_0$.
	We assume that $\theta_0$ is unique, and $\text{sgn } h_0(\theta_0+\delta) \neq \text{sgn } h_0(\theta_0-\delta)$, i.e. the function $h_0$ has opposite signs at the two sides of $\theta_0$. By the continuity of $g(x,t)$ in $t$ for any $x$, the function $h_0$ is continuous. Then, for $\epsilon$ small enough, we can define
	\begin{align*}
		C_a(\epsilon)&=\lbrace t \in \Theta : h_0(t) = \epsilon \rbrace, \quad C_b(\epsilon) =\lbrace t \in \Theta : h_0(t) = -\epsilon \rbrace,\\
		t_a(\epsilon) &= \arg\min_{t \in C_a} |t-\theta_0|, \quad t_b(\epsilon) = \arg\min_{t \in C_b} |t-\theta_0|, 
	\end{align*}
	i.e. $t_a$, $t_b$ are the closest points to $\theta_0$ with $|h_0(t)|$ equal to $\epsilon$. Due to continuity of $h_0(t)$, $\epsilon_1 <\epsilon_2 \implies |t_a(\epsilon_1)-\theta_0| \leq |t_a(\epsilon_2)-\theta_0|$, similarly for $t_b$, so that these points can only get closer to $\theta_0$ when we decrease the deviation $\epsilon$. Define 
	\begin{align*}
		&U_{t_a,t_b}(\epsilon)= \\
		&\left\lbrace f \in \mathcal{F} : \left|\int g(x,t_a(\epsilon))(f(x)-f_0(x))\mathrm{d}x\right| <\epsilon^c ,\left|\int g(x,t_b(\epsilon))(f(x)-f_0(x))\mathrm{d}x\right| <\epsilon^c \right\rbrace,
	\end{align*} which are weak neighbourhoods of $f_0$ indexed by $\epsilon$, for arbitrary $c$. We will set $c$ according to $\epsilon$, such that when $\epsilon < 1$, $c>1$, whereas when $\epsilon \geq 1$, $c<1$. We have that
	\begin{align*}
		&\epsilon<1, c>1, \text{ and } f\in U_{t_a,t_b}(\epsilon) \text{ OR } \epsilon>1 ,c<1,\text{ and } f\in U_{t_a,t_b}(\epsilon)\\
		&\implies \int g(x,t_a(\epsilon))f(x)\mathrm{d}x \in (\epsilon-\epsilon^c, \epsilon+\epsilon^c) >0 \text{ and }\\
		&\quad\quad \int g(x,t_b(\epsilon))f(x)\mathrm{d}x \in (-\epsilon-\epsilon^c, -\epsilon+\epsilon^c)<0\\
		&\implies \theta(f)\in\left( \min\left(t_a(\epsilon), t_b(\epsilon)\right), \max\left(t_a(\epsilon), t_b(\epsilon)\right)\right)\\
		&\implies |\theta(f)-\theta_0|< \max(|t_a(\epsilon)-\theta_0|,|t_b(\epsilon)-\theta_0|),
	\end{align*}
	the second to last implication is due to continuity of $g(x,t)$ in $t$ for any $x$ hence that of $\int g(x,t) f(x)\mathrm{d}x$. Define $e(\epsilon) = \max(|t_a(\epsilon)-\theta_0|,|t_b(\epsilon)-\theta_0|).$ Hence
	$
	U_{t_a,t_b}(\epsilon) \subseteq \lbrace f\in \mathcal{F}: |\theta(f)-\theta_0|< e(\epsilon) \rbrace
	$.
	Recall that both $t_a(\epsilon)$ and $t_b(\epsilon)$ $\rightarrow \theta_0$ as $\epsilon\rightarrow0$ (monotonically), such that $e(\epsilon)\rightarrow 0$ monotonically as $\epsilon \rightarrow 0$.
	
	Then, given weak consistency of the proposal model $\mathcal{P}_\Pi$ with measure $\Pi$, $\forall \epsilon>0$, we have,
	\begin{align*}
		\lim_{n\rightarrow \infty} \Pi\left(U_{t_a,t_b}(\epsilon) \mid X_1,X_2,\ldots,X_n \right) = 1\label{eq:weakNeibourhood_ta_tb}\\
		\implies \lim \Pi(|\theta(f)-\theta_0)|< e(\epsilon) \mid X_1,X_2,\ldots,X_n) = 1.
	\end{align*} 
	
	Hence, posterior parametric inference for a parameter defined via estimating equation $g(x,t)$ that is continuous in $t$ and integrable will be consistent if the proposal nonparametric model is at least weakly consistent for the true data-generating distribution $f_0$. The result above easily extends to the case where the data-generating distribution is discrete or a mixed distribution, by considering the definition of weak neighbourhoods of distribution functions which appears on p. 81 of \cite{ghosh2003bayesian}.
	
	\subsection*{Parametric consistency of $\theta$-augmented Bayesian posterior for parameters defined via estimating equations when the proposal model is the Dirichlet process}
	In the case that $\mathcal{P}_\Pi$ is the Dirichlet process, we have that for functional parameters defined via an estimating equations $g(x,t)$, parametric consistency does not require a boundedness condition on $g(x,t)$. We have that continuity of $g(x,t)$ in $x$ and $t$, integrability of $\int g(x,t) \mathrm{d}F(x)$ for all $F$ in the support of $\Pi$, and integrability of $\int g(x,t) \mathrm{d}F_0(x)$ at any given $t$ together imply the parametric consistency of $\mathcal{P}_\Pi$.
	
	To see this, the proof contained in the previous section can be used, with the only modification being the definition of the neighbourhood $U_{t_a,t_b}(\epsilon)$ as
	\begin{align*}
		&U_{t_a,t_b}(\epsilon)=\\
		&\left\lbrace F \in \mathcal{F} : \left|\int g(x,t_a(\epsilon))\mathrm{d} (F(x)-F_0(x))\right| <\epsilon^c ,\right.
		\left.\left|\int g(x,t_b(\epsilon))\mathrm{d} (F(x)-F_0(x))\right| <\epsilon^c \right\rbrace.
	\end{align*} 
	Let the prior DP under the proposal model be denoted by $\text{DP}(\phi, G_0)$. We take note of Theorem 4.16 of \cite{ghosal2017fundamentals}, the final assertion of which implies that, at every fixed $t$, $\int g(x,t) \mathrm{d}F_0(x)$ is the limit of the random quantity $\int g(x,t) \mathrm{d}F(x)$, where $F\sim \text{DP} \left(\phi+n,(\frac{\phi}{\phi+n})G_0 + (\frac{n}{\phi+n})F_n\right)$.
	As this limit is a constant, Proposition 4.3 of \cite{ghosal2017fundamentals} implies that
	\begin{align*}
		1 = \lim_{n\rightarrow\infty} \Pi\left(\left|\int g(x,t_a(\epsilon))\mathrm{d} (F(x)-F_0(x))\right| \leq \epsilon^c \mid X_1\ldots,X_n\right) \leq \\
		\lim_{n\rightarrow\infty} \Pi \left(U_{t_a,t_b}(\epsilon)\mid X_1,\ldots,X_n\right),
	\end{align*}
	$\forall \epsilon>0$ at arbitrary $c$, as long as $g(x,t)$ is a integrable function at any fixed $t$.
	
	\section{Details regarding Section \ref{sec:compare-meanvar}}\label{sec:SupplmentDetails_meanvar}
	\subsection*{Introduction of competing methods}
	We give some details to the alternative Bayesian semiparametric methods under comparison in Section \ref{sec:compare-meanvar}. We wish to obtain posterior inference conditional on some dataset $\tilde{x}_n = (x_1,\ldots, x_n)$. As usual, the target functional is denoted by $\theta$.
	
	Let $F_X(\cdot;\tilde{w}_n,\tilde{x}_n)$ denote a discrete distribution function for observable $X$ that is parameterized by a weight vector $\tilde{w}_n= (w_1,\ldots, w_n)$ with $w_i\geq 0$ assigned to observed data point $x_i$, $i=1,\ldots,n$, and $\sum_{i=1}^n w_i=1$; that is, $$F_X(\cdot;\tilde{w}_n, \tilde{x}_n)=\sum_{i=1}^n w_i \mathds{1}_{(-\infty, x_i]}(\cdot).$$
	The Bayesian bootstrap posterior \citep{rubin1981bayesian} is a distribution supported on random $F_X(\cdot;\tilde{w}_n,\tilde{x}_n)$ conditional on the observed $\tilde{x}_n$, of which the weight vector is \textit{a-posteriori} distributed as
	\begin{align*}
		(w_1,\ldots, w_n) \mid x_1,\ldots,x_n \sim \text{Dirichlet}(1,\ldots,1).\nonumber
	\end{align*}
	This posterior over random distributions for the observable induces a probability distribution for the functional parameter $\theta$, which we take to be Bayesian bootstrap posterior distribution for $\theta$. The method easily generalizes to $\theta=(\mu,\sigma^2)$.
	
	The Bayesian empirical likelihood \citep{lazar03} method substitutes the profile empirical likelihood function $R_n(t;\tilde{x}_n)$ \citep{owen1990empirical} for the true likelihood function in the usual Bayesian calculations, where
	\begin{align*}
		R_n(t;\tilde{x}_n) = \max_{\tilde{w}_n \in \mathcal{S}^{n-1}} \left\lbrace \prod_{i=1}^n w_i: \theta(F_X(\cdot;\tilde{w}_n, \tilde{x}_n))=t \right \rbrace.
	\end{align*}
	Given a subjective prior $p_\theta(t)$ for the target parameter, the Bayesian empirical likelihood posterior is
	$
	\pi_{\text{BEL}}(\theta|\tilde{x}_n) \propto R_n(\theta;\tilde{x}_n) p_\theta(\theta).
	$
	In our simulation study, to infer $\theta=(\mu,\sigma^2)$, $R_n$ was defined as
	\begin{align*}
		R_n(\mu,\sigma^2;\tilde{x}_n) := \max_{\tilde{w}_n\in \mathcal{S}^{n-1}} \left\lbrace \prod_{i=1}^n w_i: \sum_{i=1}^n \left( x_i - \mu\right) = 0 \text{ and } \sum_{i=1}^n \dfrac{1}{n}(x_i -\mu)^2 - \sigma^2 = 0 \right \rbrace,
	\end{align*}
	and Bayesian empirical likelihood posterior as proportional to $R_n(\mu,\sigma^2;\tilde{x}_n) p_{\mu,\sigma^2}(\mu,\sigma^2)$.
	
	The general Bayes (GB) method of \cite{bissiri2016general} is used for estimating a functional parameter that is defined as the minimizer of the expectation of a loss function $l(x,t)$. This loss function has the property that under the true data-generating distribution $F_0$, the true value $\theta_0$ is identified by
	\begin{equation}
		\theta_0 = \argmin_{t \in \Theta} \int c \cdot l(x,t) \mathrm{d}F_0(x), \label{eq:identification_criteria}
	\end{equation}
	given arbitrary constant $c$.
	Suitable loss functions are typically selected to correspond to the loss functions of well-known consistent m-estimators.
	Given constant $c$ and dataset $\tilde{x}_n$, the general Bayes method proposes the conditional density function
	\begin{equation}
		\pi_{\text{GB}}(\theta \mid \tilde{x}_n) \propto \exp\left(-c\sum_{i=1}^n l(x_i,\theta)\right) p_\theta(\theta)\label{eq:GB-posteriro}
	\end{equation}
	as the optimal choice for describing belief regarding $\theta$, for reasons detailed in \cite{bissiri2016general}. 
	
	In the context of $\theta=(\mu,\sigma^2)$, we note that there does not exist any distribution-free M-estimator for the variance parameter directly, and therefore consider the joint estimation of the first two moments instead. We note that the loss functions
	\begin{align*}
		l_1(x,m_1) = (x- m_1)^2\\
		l_2(x, m_2) = (x^2 - m_2)^2
	\end{align*}
	separately identifies the first moment, $\mu$, and the second raw moment, $\mu'_{2}$. However, M-estimation for joint inference of the first and second moments is unusual, and we found ourselves modifying the original GB posterior in Eqn. (\ref{eq:GB-posteriro}) to produce efficient estimation of $(\mu,\mu'_2)$ in our simulation study. Specifically, we modified Eqn. (\ref{eq:GB-posteriro}) as
	\begin{align}
		\pi_{\text{GB}}(\mu, \mu'_2\mid\tilde{x}_n) \propto 
		\exp\left(- \sum_{i=1}^n l^{1/2}(x_i,\mu,\mu'_2)^T\times C \times l^{1/2}(x_i, \mu,\mu'_2) \right) p_{\mu,\mu'_2}(\mu, \mu'_2),\label{eq:GB-multi-dim-posteriro}
	\end{align}
	where
	\[
	l^{1/2}(x,m_1,m_2)= \begin{bmatrix}  x- m_1\\ x^2 - m_2 \end{bmatrix},
	\]
	and $C$ a $2\times2$ matrix. The chosen loss function for joint inference on $(\mu,\mu'_2)$ is therefore $l^{1/2}(x_i,m_1,m_2)^T\times C \times l^{1/2}(x_i, m_1,m_2)$, which satisfies the identification criteria as shown in Eqn. (\ref{eq:identification_criteria}).

	\subsection*{Simulation setup}
\begin{figure}
		\centering
		\subfloat[$X$]{\animage{0.45}{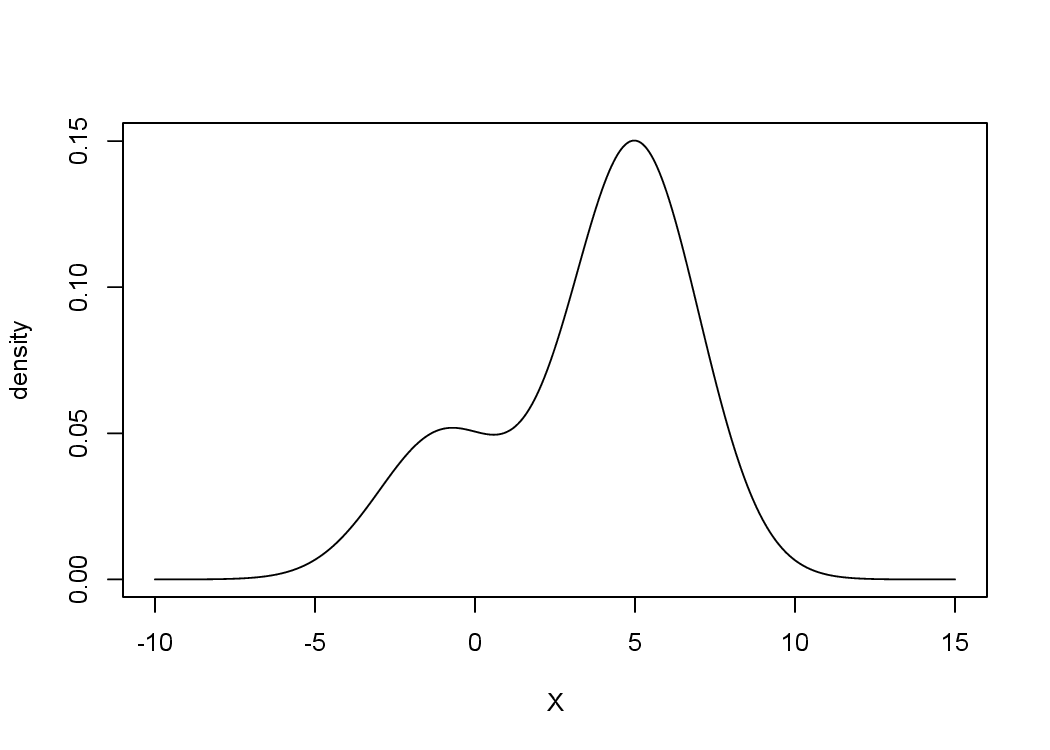}}
		\subfloat[$X^2$]{\animage{0.45}{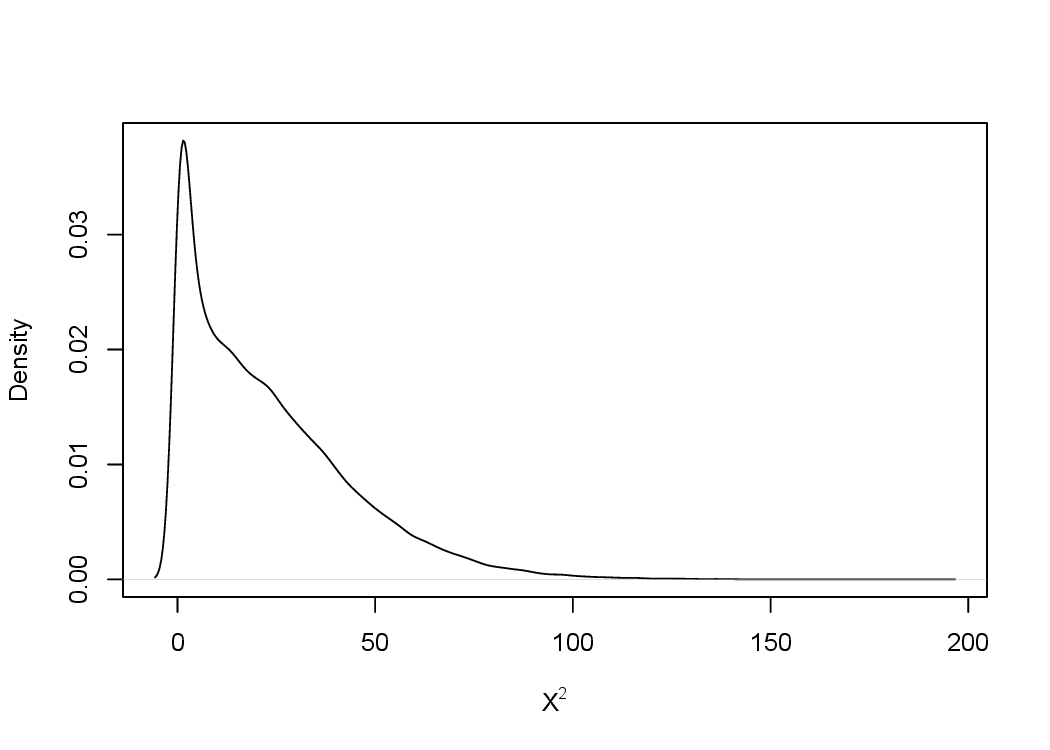}}
		\caption{Marginal density functions of random observable $X$ and $X^2$ generated by Model \ref{model:comparison-meanVAr-datagen}. The density function of $X$ is shown on the left-hand-side, and that of $X^2$ is shown on the right-hand-side.}\label{fig:data-gen-comparisom-meanvar}
	\end{figure}

	We generated the data according to Model \ref{model:comparison-meanVAr-datagen}, which is a skewed distribution. The density function for observable $X$ under Model \ref{model:comparison-meanVAr-datagen} is shown in Figure \ref{fig:data-gen-comparisom-meanvar}.
	\begin{model}\label{model:comparison-meanVAr-datagen}
		(Data-generating mechanism)
		\begin{align*}
			X &= -6 Z + T\\
			T &\sim \text{Normal}(\mu= 5, \sigma^2= 4 )\\
			Z &\sim \text{Bernoulli}(0.25).
		\end{align*}	
	\end{model}

\begin{figure}
		\centering
		\subfloat[$\mu$]{
			\animage{0.45}{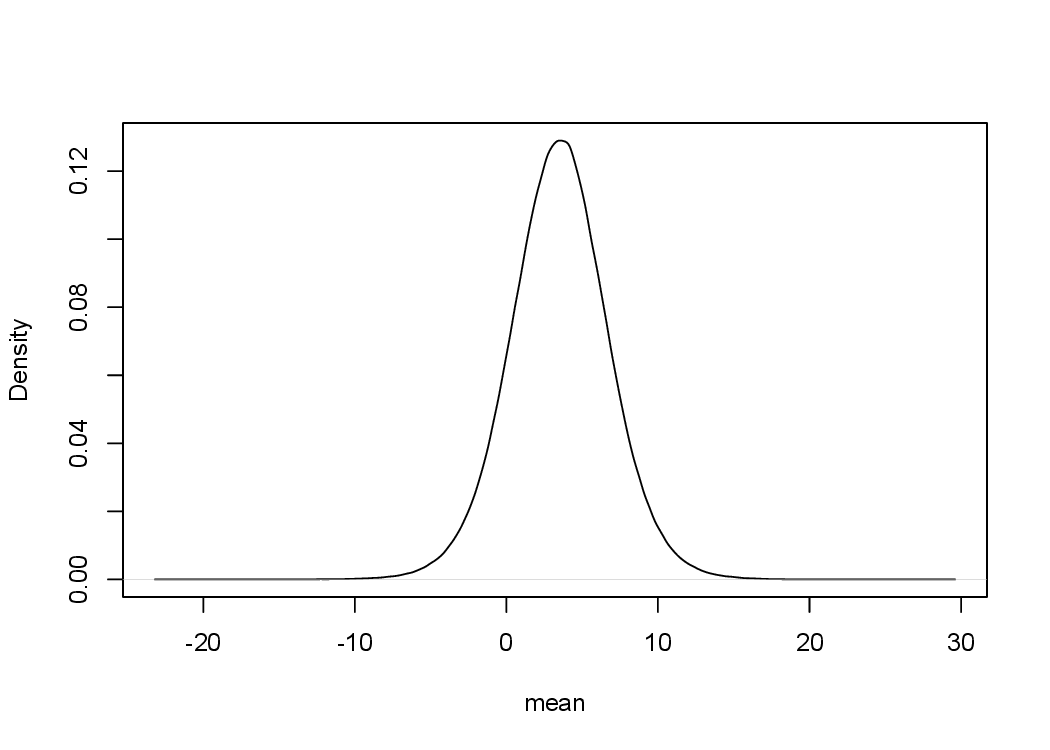}} 
		\subfloat[$\sigma^2$]{
			\animage{0.45}{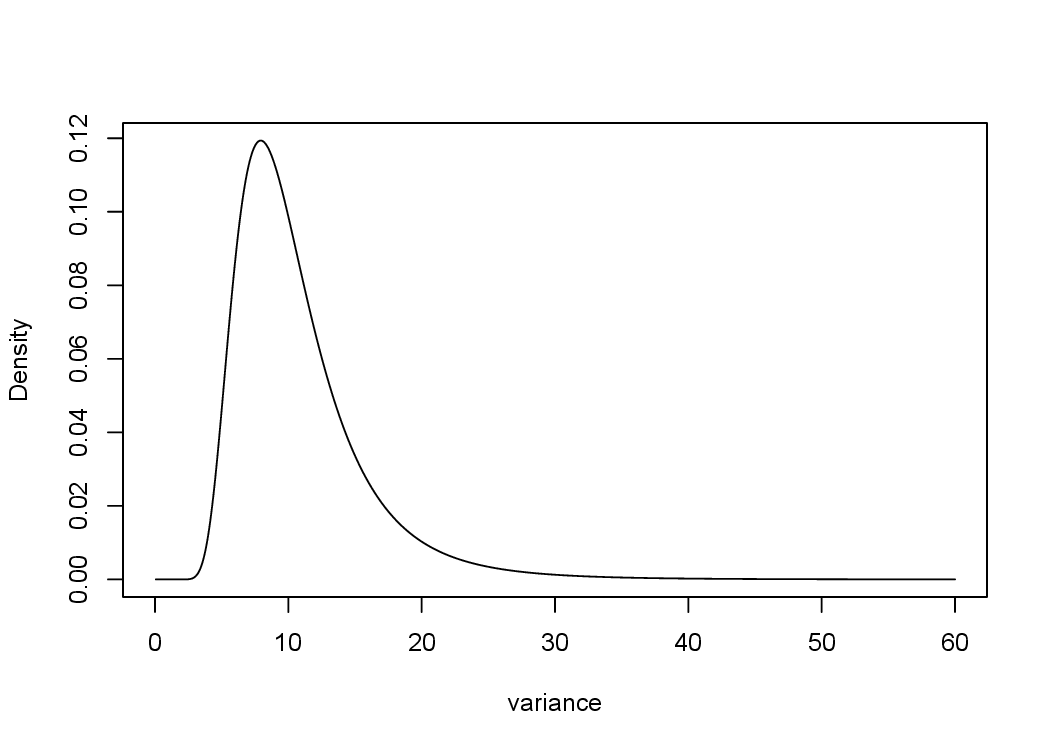}} 
		\caption{Marginal density functions of subjective prior according to Model \ref{model:comparison-meanVAr-subjectivePrior}. The marginal subjective prior for $\mu$ is shown on the left, and for $\sigma^2$ on the right.}\label{fig:margPrior-comparisom-meanvar}
	\end{figure}

	For methods under evaluation that require a subjective prior distribution, the same prior was utilized. This prior was a Normal-Inverse-Gamma (NIG) distribution with the marginal prior mean of variance parameter matching the variance of data-generating distribution, and the marginal prior mean of mean parameter matching the mean of data-generating distribution;  see Model \ref{model:comparison-meanVAr-subjectivePrior}. The densities of the marginal prior distributions are shown in Figure \ref{fig:margPrior-comparisom-meanvar}.
	\begin{model}\label{model:comparison-meanVAr-subjectivePrior}
		(Joint subjective prior  $p_{\mu,\sigma^2}(\mu, \sigma^2)$ in Section \ref{sec:compare-meanvar})
		\begin{align*}
			1/\sigma^2 &\sim \text{Gamma}(\alpha= 6.623, \beta= 60.442)\\
			\mu \mid \sigma^2 &\sim \text{Normal}(\mu_0=3.5, \sigma_0^2=\sigma^2 ). 
		\end{align*}	
	\end{model}

	To implement $\theta$-augmented Bayesian (TAB) inference, we assumed the Bayesian prior model
	\begin{align*}
		X &\sim F_X\\
		F_X &\sim \TA(m= p_{\mu,\sigma^2}/q^\Pi_{\mu, \sigma^2}, \mathcal{P}_\Pi= \text{Model \ref{model:comparison-meanVAr-TAB-proposal}})
	\end{align*} 
	for the observable, $q^\Pi_{\mu, \sigma^2}$ being the density function of $(\mu, \sigma^2)$ induced by $\mathcal{P}_\Pi$ \textit{a priori}. The Dirichlet process model detailed in Model \ref{model:comparison-meanVAr-TAB-proposal} was the chosen proposal model for this TA model.
	
	\begin{model}\label{model:comparison-meanVAr-TAB-proposal}
		(Proposal model for the $\theta$-augmented inference in Section \ref{sec:compare-meanvar})
		\begin{align*}
			F_X\sim& \text{DP}(\phi, G_0')\\
			G_0' =& \text{ Discretized version of continuous distribution $G_0$,}\\
			&\text{ with mass assigned to points $\lbrace{ ih\mid  i\in \mathds{Z} \rbrace}$ for fixed bin width $h$ s.t.}\\
			& G_0'(X=ih) = \int\mathds{1} (2i-1)h/2 \leq x \leq (2i+1)h/2] \mathrm{d}G_0(x)\\
			G_0 &= \text{Normal}(\mu_0=0,\sigma_0^2=100)\\
			\phi&=0.5\\
			h&=1\text{e}-5.
		\end{align*}	
	\end{model}

\begin{figure}
		\centering
		\animage{0.75}{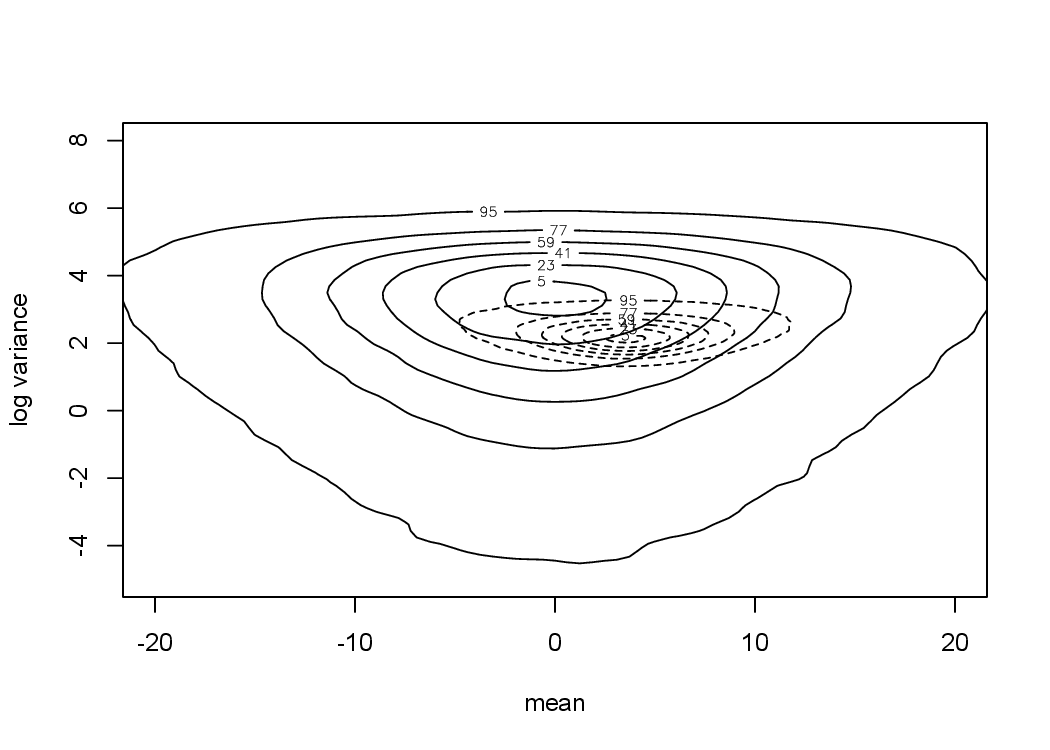}
		\caption{Contour plots comparing $q^\Pi_{\mu,\sigma^2}$, in solid lines, versus the $p_{\mu,\sigma^2}$, in dotted lines.  In this plot, distributions are parameterized in terms of mean and log variance.}\label{fig:comparison-meanvar-ptoq}
	\end{figure}

\begin{figure}
		\centering
		\animage{0.75}{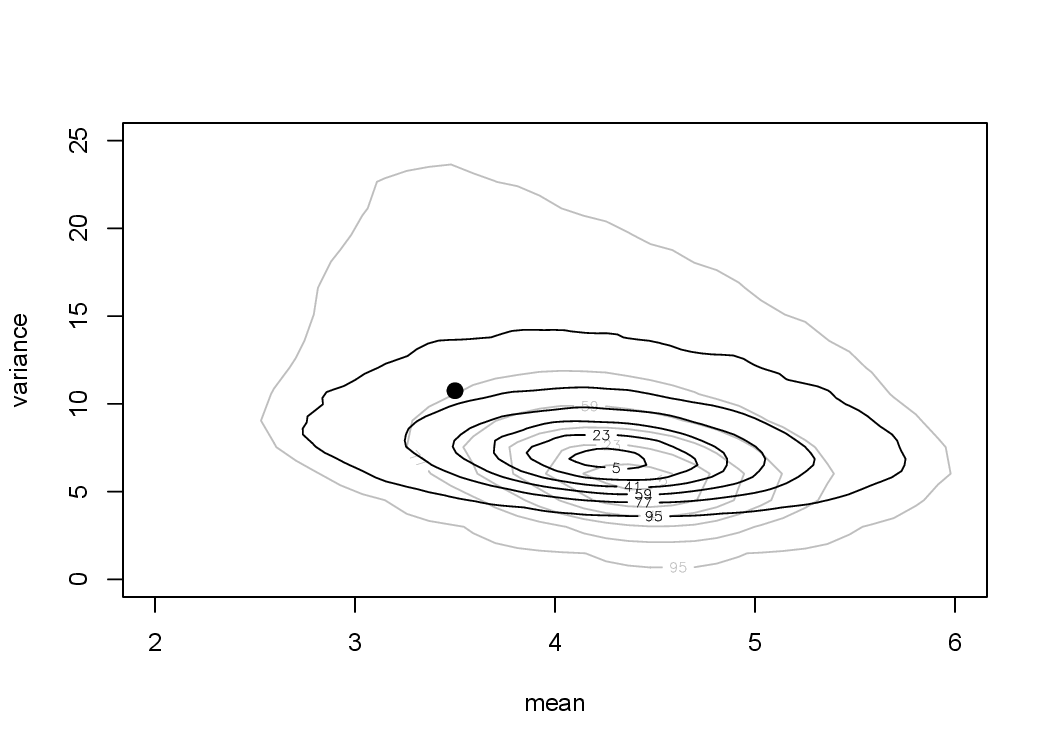}
		\caption{Contour plot of a posterior distribution under the $\theta$-augmented model, in black, and that of the corresponding posterior under the proposal model, in grey, based on a particular dataset with 20 samples. The contours are shown under $(\mu, \sigma^2)$ parameterization because the posterior densities do not concentrate around 0 for $\sigma^2$ to necessitate a conversion to log variance scale. }\label{fig:comparison-meanvar-TABresult}
	\end{figure}

	In the simulation study, TAB inference was obtained via Markov chain Monte Carlo (MCMC) according to the method outlined in Section \ref{sec:algo_MCMC} with $4\times10^5$ runs for each dataset. We note that the parametric density function induced by the proposal \textit{a priori}, $q^{\Pi}_{\mu,\sigma^2}$, is much more disperse than the subjective prior; see Figure \ref{fig:comparison-meanvar-ptoq}. The parametric posterior induced by the proposal model \textit{a posteriori}, $q^{\Pi}_{\mu,\sigma^2 \mid \tilde{x}_n}$, is also much more concentrated than $q^{\Pi}_{\mu,\sigma^2}$ for the generated datasets.  Therefore substitution of estimated $\hat{q}^{\Pi}_{\mu,\sigma^2}$ for $q^{\Pi}_{\mu,\sigma^2}$ in the MCMC provided a good approximation to the exact TAB posterior under the $\theta$-augmented model; see Figure \ref{fig:comparison-meanvar-TABresult} for an example. Marginal inference on $\mu$ and $\sigma^2$ were obtained by marginalization of the joint posterior.
	
	
	The general Bayes posterior for $(\mu,\mu'_2)$, i.e. Eqn. (\ref{eq:GB-multi-dim-posteriro}), was obtained for each simulated dataset via the MCMC approximation, with a Markov chain length of $1\times 10^5$ for each dataset. For each Markov chain, the tuning matrix, $C$, was chosen based on the dataset; specifically it was set 1/2 times the inverse of sample variance covariance matrix of $(x, x^2)$. This choice was made based on asymptotic tuning and similar in spirits as Section 3.2 of \cite{bissiri2016general}. As for $p_{\mu,\mu'_2}(\mu, \mu'_2)$, it was obtained through reparametrization of $p_{\mu,\sigma^2}(\mu,\sigma^2)$ via the change of variable formula. For a given dataset $\tilde{x}_n$, the corresponding MCMC proposal distribution was a multivariate Gaussian distribution with mean of $(\bar{x}_n, \frac{1}{n}\sum_{i=1}^{n}x_i^2)$ and variance of $C/n$, truncated to valid values of $(\mu, \mu'_2)$ satisfying $\mu'_2 - \mu^2 > 0$. MCMC samples of $(\mu, \mu'_2)$ were subsequently transformed to $(\mu, \sigma^2)$ to approximate $\pi_{\text{GB}}(\mu, \sigma^2\mid\tilde{x}_n)$, $\pi_{\text{GB}}(\mu\mid\tilde{x}_n)$ and $\pi_{\text{GB}}(\sigma^2\mid\tilde{x}_n)$.

\begin{figure}
		\centering
		\animage{0.75}{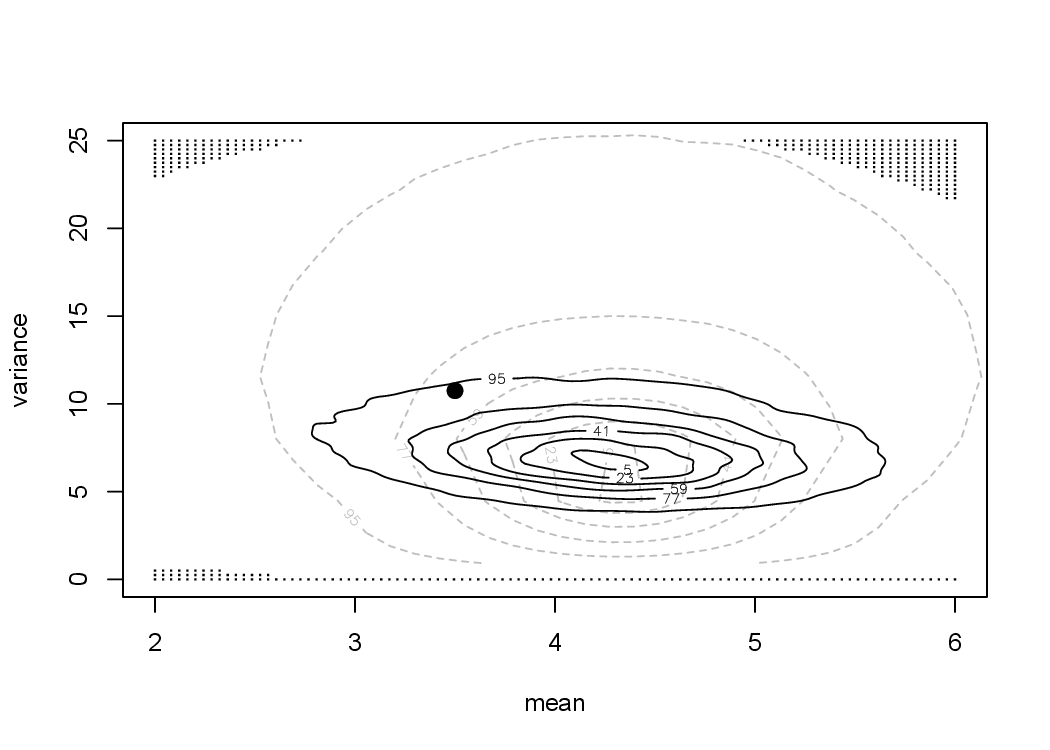}
		\caption{Contour plot of a Bayesian empirical likelihood posterior distribution (solid lines) and that of the corresponding Bayesian empirical likelihood MCMC proposal distribution (in grey dotted lines) based on a particular dataset. Square dots mark the locations (on a grid) outside of the valid parameter space as given by the convex hull condition. The figure shows the contours under $(\mu,\sigma^2)$ parametrization since the Bayesian empirical likelihood posterior in question is not concentrated around 0 for $\sigma^2$ to necessitate a conversion to log variance scale. }\label{fig:comparison-meanvar-BEL-propvsPost}
	\end{figure}

	Finally, to implement Bayesian empirical likelihood inference, the profile empirical likelihood function $R_n$ was obtained via the iterative least squares technique described in Section 3.14 of \cite{owen2001empirical}. Due to a lack of explicit formula for $R_n$, finding good proposal distribution for the MCMC approximation of a Bayesian empirical likelihood posterior requires trial-and-error. On the first pass, we chose the MCMC proposal distribution based on the posterior NIG distribution that would have resulted from a misspecified normal likelihood with a conjugate NIG prior equivalent to our subjective prior, while scaling the shape and rate parameters by 0.25 to ensure that the proposal distribution was disperse enough.  This method of setting the proposal distribution worked well most of the time; we show a plot comparing the default MCMC proposal for Bayesian empirical likelihood inference to the target Bayesian empirical likelihood posterior of a particular dataset in Figure \ref{fig:comparison-meanvar-BEL-propvsPost}. Due to the low dimensionality of the target parameter, it was feasible to visually check for the quality of the MCMC proposal distribution. When the aforementioned MCMC proposal did not work well, it was adjusted until it became wider than the Bayesian empirical likelihood posterior. Due to the time intensive nature of the iterative least squares algorithm for finding $R_n$ at given values of $(\mu, \sigma^2)$, each MCMC chain contained just $1\times 10^4$ runs.
	
	Performance of the four competing methods were explored at two sample sizes $(n=20, n=50)$. At each sample size level, $300$ datasets were generated, and the performance metrics were calculated based on the collection of posterior distributions conditional on each dataset. We calculated the coverage probability and expected size of 95\% highest posterior density  region for joint and marginal inference. For marginal inference, we also calculated the absolute bias of the Bayes estimator, which is the mean of a Bayesian posterior, and the expected quadratic risk, ${E}_{F_0}[\int(\mu - \mu_0 )^2 \pi(\mathrm{d}\mu\mid \tilde{x}_n)]$, ${E}_{F_0}[\int(\sigma^2 - \sigma^2_0)^2 \pi(\mathrm{d}\sigma^2 \mid \tilde{x}_n)]$. Results from the simulation study is summarized in Table~\ref{tab:compare-meanvar_2d} for joint inference and Table \ref{tab:compare-meanvar_mean}, Table \ref{tab:compare-meanvar_var} for marginal inference.

	\section{Details regarding Section \ref{sec:MAR}}\label{append:Ex2}
	In Section~\ref{sec:MAR}, we simulated the data from four data-generating mechanisms, given in Model \ref{model:MAR-LogisticRegression}- \ref{model:MAR-notLogisticRegression-nonlinear}. We conducted Bayesian inference with a $\theta$-augmented model with the proposal model being Model \ref{model:tabProp-MAR}.
	
	\begin{model}\label{model:MAR-LogisticRegression}
		(Data Generating Distribution 1)
		\begin{align*}
			&\beta_1 = 0.5; \beta_0 = 1\\
			&\sigma^2_\epsilon = 4\\
			&X \sim \text{Normal}(10, 100)\\
			&e \sim \text{Normal}(0, 4 )\\
			&Y\mid X = 1 + 0.5 X + e\\
			&C\mid X \sim \text{Bernoulli}\left(p= \left(1+\exp\left(-(X-10)/10\right)\right)^{-1}\right)
		\end{align*}
	\end{model}
	
	\begin{model}\label{model:MAR-LogisticRegression-wronglinear}
		(Data Generating Distribution 2)
		\begin{align*}
			&X \sim \text{Normal}(10, 100)\\
			&e \sim \text{Normal}(0, 4 )\\
			&Y\mid X = 0.006(X^2 + 40 X + 400) + e\\
			&C\mid X \sim \text{Bernoulli}\left(p= \left(1+\exp\left(-(X-10)/10\right)\right)^{-1}\right)
		\end{align*}
	\end{model}
	
	\begin{model}\label{model:MAR-notLogisticRegression-linear}
		(Data Generating Distribution 3)
		\begin{align*}
			&\beta_1 = 0.5; \beta_0 = 1\\
			&X \sim \text{Normal}(10, 100)\\
			&e \sim \text{Normal}(0, 4 )\\
			&Y\mid X = 1 + 0.5 X + e\\
			&C\mid X \sim \text{Bernoulli}\left(p= \Phi((X-10)/10)\right).
		\end{align*}
	\end{model}
	
	\begin{model}\label{model:MAR-notLogisticRegression-nonlinear}
		(Data Generating Distribution 4)
		\begin{align*}
			&\beta_1 = 0.5; \beta_0 = 1\\
			&X \sim \text{Normal}(10, 100)\\
			&e \sim \text{Normal}(0, 4 )\\
			&Y\mid X = 0.006(X^2 + 40 X + 400)+ e\\
			&C\mid X \sim \text{Bernoulli}\left(p= \Phi((X-10)/10)\right).
		\end{align*}
	\end{model}
	
	\begin{model}\label{model:tabProp-MAR}
		(Proposal model for estimating the mean with data missing at random)
		\begin{align*}
			F_{X, C, CY}&\sim \text{DP}(\phi, G_0')\\
			\phi&=0.5\\
			h &= 1e-4\\
			G_0'& = \text{ Discretized version of $G_{0X}\times G_{0C}\times G_{0CY}(\cdot\mid C)$,}\\
			&\text{ with mass assigned to points $\lbrace{ (ih,c,jh) \mid  i,j\in \mathds{Z} \rbrace}$ for fixed bin width $h$ s.t.}\\
			&G_0'(X=ih, C= c, CY=jh\mid h) = G_{0C}(c)\times \\
			&\quad\quad\quad\quad\quad\quad\quad\quad\quad\quad\int \int\mathds{1}[ (2i-1)h/2 \leq x \leq (2i+1)h/2]\times\\
			&\quad\quad\quad\quad\quad\quad\quad\quad\quad\quad\mathds{1} [(2j-1)h/2 \leq cy \leq (2j+1)h/2] \mathrm{d}G_{0X}(x)\mathrm{d}G_{0CY}(y\mid c)\\
			G_{0X} &= \text{Normal}(\mu=10, \sigma^2=1)\\
			G_{0C} &= \text{Bernoulli}(0.2)\\
			&G_{0CY}(\cdot\mid C=1) = \text{Normal}(\mu=0, \sigma^2=50^2)\\
			&G_{0CY}(\cdot\mid C=0) = 0 \text{ with probability 1}.
		\end{align*}
	\end{model}

	There is a computational limit on the precision parameter of the proposal Dirichlet process model, such that while we know as $\phi$ tends to 0 the posterior distribution under the proposal model better approximates the Bayesian bootstrap, the prior distribution under the proposal model becomes harder and harder to sample from. Here, the results were obtained with the $\phi$ parameter of the proposal model set to 0.5, which has worked well in past experience. Several considerations in particular influenced our choice of a base distribution for the proposal model. Firstly, we note that the functionals of interest involves weighting of the residuals of regression $(Y-{E}[Y\mid X])$ by the inverse of $\hat{p}(C=1;X)$. As such, the target functional is extremely sensitive to extreme values of $X$ is in the support of the distribution for the observables. When specifying the proposal model, we chose a base distribution for $X$ that does not place much weight in the extreme values of observed $X$, in order to have the effective likelihood, $q^\Pi_{\theta\mid \tilde{x}_n}/q^\Pi_\theta$, resemble the Bayesian bootstrap, which is known to have good asymptotic properties. Secondly, we decided that the proposal model base distribution for $CY$ should be sufficiently wide to reflect our relative lack of information in $Y$ \textit{a priori}. The base distribution for $C$ curiously plays a role in terms of how much the prior base distribution contributes to the posterior distribution of the functional- if chance of observing $Y$ is small in the base distribution, then the proposal posterior will be more data driven. In the simulation each posterior distribution under the $\theta$-augmented model was obtained via an MCMC chain with $1\times 10^5$ runs. 
	
	The Frequentist metrics for evaluating the resulting posterior inference are the same as those of Section \ref{sec:compare-meanvar}, and are detailed in Appendix \ref{sec:SupplmentDetails_meanvar}. These measures of performance were estimated with a minimum of 500 datasets from each data-generating mechanism.

	\bibliography{../../../bibliography}

\end{document}